\documentclass[nopreprintline,12pt,authoryear]{elsarticle}
\usepackage{amssymb}
\usepackage{lineno}
\usepackage{longtable}
\usepackage{fullpage}
\usepackage{amsmath}
\usepackage{color,hyperref,graphicx,pdfsync,overpic,color,epstopdf,rotating,dashrule,float,bm,multicol,multirow,setspace}
\usepackage{tabularx,booktabs}
 \usepackage{subfig}
\usepackage{mathrsfs}
\usepackage{mathtools}
\usepackage{textcomp}
\usepackage{setspace}
\usepackage{listings}
\usepackage{array}
\usepackage{comment}
\usepackage{enumitem}

\usepackage[dvipsnames,table]{xcolor}
\usepackage{enumitem} %remove left margin in enumerate
\usepackage{wrapfig}

\newcommand\Rey{\mbox{\textit{Re}}}  % Reynolds number
\newsavebox{\astrutbox}
\sbox{\astrutbox}{\rule[-5pt]{0pt}{20pt}}

\newcommand{\defeq}{\vcentcolon=}

\newcommand{\mat}[1]{{\bm{#1}}}

\newcommand{\mA}{\mat{A}}

\newcommand{\mD}{\mat{D}}

\newcommand{\mI}{\mat{I}}

\newcommand{\mQ}{\mat{Q}}

\newcommand{\sD}{\mathcal{D}}

\newcommand{\m}[1]{\boldsymbol{#1}}

\newcommand{\bc}{{ \bf c}}

\newcommand{\bbf}{{ \bm{f}}}

\newcommand{\bg}{{ \bf g}}

\newcommand{\bu}{{ \bm{u}}}

\newcommand{\bq}{{ \bm q}}
\newcommand{\Laplace}{\Delta}

\journal{International Journal of Heat and Fluid Flow}

\begin{document}
%\linenumbers

\begin{frontmatter}



\title{Sparsity-promoting methods for isolating dominant linear amplification mechanisms in wall-bounded flows} %% Article title


\author[label1]{Scott T. M. Dawson} %% Author name
\author[label1]{Jaime Prado Zayas} %% Author name
\author[label1]{Barbara Lopez-Doriga} %% Author name

\affiliation[label1]{organization={Mechanical, Materials, and Aerospace Engineering Department, Illinois Institute of Technology},%Department and Organization
            %addressline={10 W 32nd St}, 
            %city={Chicago},
            %postcode={60616}, 
            %state={Illinois},
            %country={USA}
            }

%% Abstract
\begin{abstract}
This work proposes a method to identify and isolate the physical mechanisms that are responsible for linear energy amplification in fluid flows. 
This is achieved by applying a  sparsity-promoting methodology to the resolvent form of the governing equations, solving an optimization problem that balances retaining the amplification properties of the original operator with minimizing the number of terms retained in the simplified sparse model.  This results in simplified operators that often have very similar pseudospectral properties as the original equations. 
 The method is demonstrated on both incompressible and compressible wall-bounded parallel shear flows, where the results obtained from the proposed method appear to be consistent with known mechanisms and simplifying assumptions, such as the lift-up mechanism, and (for the compressible case) Morkovin's hypothesis and the strong Reynolds analogy. 
 This provides a framework for the application of this method to problems for which knowledge of pertinent amplification mechanisms is less established.
\end{abstract}

% temporarily suppress these


%% Keywords

\end{frontmatter}

\section{Introduction}
\label{sec:intro}
While turbulent fluid flows exhibit highly nonlinear dynamics, there is strong evidence that linear mechanisms play a key role in both the formation of coherent structures within and the overall statistics of many such flows.  This is particularly true for shear-driven turbulence, where spatial gradients of the mean velocity field and the non-normality of the linearized equations can drive very large linear amplification \citep{trefethen1993science,schmid2012book,hwang2010linear}. 
%mechanisms that play a major role in the structure and statistics of the turbulent flow.
 Indeed, several recent works  %\cite{abreu2020spectral,nogueira2021forcing,pickering2021optimal}) 
\citep{abreu2020spectral,tissot2021stochastic,nogueira2021forcing,pickering2021optimal,symon2023use} 
 have demonstrated agreement between coherent structure prediction via resolvent analysis of the mean-linearized equations  \citep{mckeon2010resolvent}, and the highest-energy structures identified directly from data via spectral spectral proper orthogonal decomposition \citep{towne2018spectral}.
 
 In configurations that have been comprehensively studied, %experimental, computational, and theoretical ,
  there is broad understanding of both the characteristics of coherent structures that form in turbulent flows, and the mechanisms that lead to their formulation. In canonical incompressible wall-bounded turbulent flows, the Orr \citep{orr1907stability} and lift-up \citep{landahl1975wave} mechanisms play a key role in the generation and evolution of features such as near-wall streaks \citep{kline1967structure}, %hairpin structures  \cite{theodorsen1952mechanisms,head1981new}, 
 and large- and very-large-scale motions %\citep{zhou1999mechanisms,kim1999very,guala2006large,hutchins2007evidence}. 
\citep{zhou1999mechanisms,hutchins2007evidence}. 
 In particular, the mechanisms giving rise to such structures can be understood through the action of a small number of terms within the governing equations. 
 The development of a similar level of understanding for a broader class of more complex, non-canonical geometries 
can be accelerated through methods that can automatically identify the terms within the governing equations that are primarily responsible for the dominant coherent features observed in such systems. 
% Add paragraph on compressible cases. 
In the compressible regime, similar structures are observed \citep{samimy1994streamwise,smith1995visualization,ganapathisubramani2006large}, though there can be notable Mach number effects in the quantitative characteristics of such structures \citep{smits1989comparison,duan2011direct}.

The present work develops a method to automatically extract such minimal-physics mechanisms from the governing equations.
  This is achieved by utilizing ideas from compressive sensing \citep{candes2006robust,candes2008introduction}, which allows such problems to be solved with convex methods, by formulating optimization problems involving the $l_1$ norm (which we will refer to as the 1-norm throughout). 
Such sparsity-promoting methods have previously been used for a range of modeling problems in fluid mechanics. From a data-driven perspective, sparsity-promoting methods have been utilized to identify low-dimensional reduced-order models that are sparse in the space of coefficients of terms within a predefined library \citep{brunton2016sindy, loiseau2018constrained,kaiser2018sparse,candon2024optimal}, and to identify active terms within governing differential equations \citep{callaham2021learning}. In the context of dynamic mode decomposition 
 \citep{Schmid:2010}, sparsity-promoting methods have been used to obtain a sparse set of nonzero mode coefficients \citep{jovanovic2014dmdsp}, to identify flow regimes \citep{kramer2017sparse} and Koopman-invariant subspaces \citep{pan2021sparsity}, and for the reconstruction of temporally \citep{tu2014compressed} and spatially  {\citep{Brunton2015jcd}} under-resolved data. 
Sparsity-promoting methods have also been leveraged to sparsify reduced-order models identified via Galerkin projection  
\citep{rubini2020l1,rubini2022priori}. %identification of a sparse set of active dynamic modes that best represent time-resolved data \citep{jovanovic2014dmdsp}, 
 In contrast, the present approach is largely data-free, applying sparsity promotion directly upon the governing equations.  
This work focuses on analysis of the resolvent form of the mean-linearized Navier--Stokes equations. Sparsity promotion has previously been applied in such analyses for the purposes of identifying spatially \citep{foures2013localization,skene2022sparsifying,mushtaq2024identifying} or spatio-temporally localized \citep{lopez2024sparse} structures. Here, rather than seeking sparsity in the structures corresponding to linear amplification mechanisms, we instead aim to sparsify the underlying linear operator, in order to identify the components of the operator that are primarily responsible for the leading linear energy amplification mechanisms identified through resolvent analysis. 

The paper proceeds as follows. In Section \ref{sec:methods}, we outline the theory and methodology underpinning this analysis, including both sparsity-promoting optimization methods, and the resolvent analysis framework that these methods will be applied to. This is followed by the application of these methods in Section \ref{sec:results}, where we first apply our sparsity-promoting methods to a simple differential operator, before applying it to both incompressible and compressible wall-bounded flows. Conclusions are provided in Section \ref{sec:conclusions}.

\section{Methodology}
\label{sec:methods}
This section introduces the methods that will be used throughout this work. We first describe the general approach for sparsity-promoting operator simplification in Section \ref{sec:sparse}. Following this, we introduce the resolvent analysis formulation of the Navier--Stokes equations in Section \ref{sec:resolvent}. Section \ref{sec:resSparse} demonstrates how this sparsity-promoting operator simplification method can be formulated and applied in the context of the resolvent formulation of the governing equations.

\subsection{Sparsity-promoting operator simplification}
\label{sec:sparse}
Before analyzing fluids systems, here we provide a more general formulation of the class of methods that will be utilized to perform sparsity-promoting analysis. We focus on analysis of linear operators, with the objective being to find an approximation of a known linear operator $\mA$ by another operator $\mA_a$ which possesses desired properties, which here will relate to the operator being sparse (in some sense). Mathematically, this is achieved by minimizing an objective function of general form
\begin{equation}
\label{eq:opt1}
    \mathcal{J}(\mA_a) = e(\mA - \mA_a) + \lambda s(\mA_a),
\end{equation}
where $e(\mA - \mA_a)$ quantifies the error in the approximation of $\mA$ by $\mA_a$, and  $s(\mA_a)$ measures the sparsity of $\mA_a$. The parameter $\lambda$ determines the extent to which the sparsity of $\mA_a$ is prioritized when solving this optimization problem. Starting from this general formulation, there are several choices that can be made to specify how the terms in equation \ref{eq:opt1} are quantified. Most simply, we may set
\begin{equation}
\label{eq:opt2}
    \mathcal{J}(\mA_a) = \|\mA - \mA_a\|_M + \lambda \|\text{vec}(\mA_a)\|_1,
\end{equation}
where $\|\cdot\|_M$ denotes a matrix norm, and the second term involves the element-wise one-norm of the approximating operator. Depending on the context, the chosen matrix norm could be the Frobenius norm (equivalent to $\|\text{vec}(\mA - \mA_a)\|_2$), or the operator 2-norm, which is equivalent to the leading singular value of $\mA - \mA_a$. The latter is the most appropriate choice where we want the leading singular values and vectors of $\mA_a$ to be close to those of $\mA$, which is desired for $\mA_a$ to retain the dominant energy amplification properties of $\mA$.
 The inclusion of vector 1-norm term in equation \ref{eq:opt2} promotes sparsity in the minimizing solution, $\mA_a$. This 1-norm is being used rather than the  0-pseudo-norm (i.e. penalizing the number of nonzero coefficients) so that the optimization problem is convex, and tractable to be solved using  standard convex optimization methods. To solve all such optimization problems in this work, we utilize SDPT3 \citep{toh1999sdpt3,tutuncu2003solving}, a primal-dual infeasible-interior-point algorithm, within the CVX convex optimization package \citep{grant2014cvx,gb08} in Matlab. 
  {
 The computational aspects of this optimization are discussed in section \ref{sec:computational}, though 
 %This method recasts equation \ref{eq:opt2} as a semi-definite program
 }we refer readers to these aforementioned references for further algorithmic details. 
After optimizing equation \ref{eq:opt2} using iterative methods, we typically set entries of $\mA_a$ below a certain threshold (typically $10^{-6}$) to zero. In subsequent sections, we will describe several variants of this general optimization approach, including several choices for the matrix norm, and constraining the entries of $\mA_a$ to preserve the blockwise structure of $\mA$. 

\subsection{Resolvent analysis}
\label{sec:resolvent}
We start by considering a general nonlinear system of differential equations of the form 
\begin{equation}
\label{eq:g}
\frac{\partial \bu}{\partial t} = \bg(\bu),
\end{equation}
where $\bu(t)$ is the system state, and $\bg$ is a nonlinear function (that can include spatial derivatives). Expanding the state about a nominal average configuration $\bu(t) = \bu_0 +  {\bu'(t)}$, equation \ref{eq:g} can be written as
\begin{equation}
\label{eq:gl}
\frac{\partial \bu'}{\partial t} = \bg(\bu_0+\bu') = \left.\frac{\partial \bg}{\partial \bu}\right|_{\bu_0}\bu'+ \bbf(\bu),
\end{equation}
where $\bbf$ accounts for the nonlinear terms remaining after the linearized dynamics are accounted for. Considering temporal Fourier modes of the form 
\begin{align}
\bu' &= \hat \bu \exp(-i\omega t), \\
\bbf &= \hat \bbf \exp(-i\omega t),
\end{align}
equation \ref{eq:gl} can be rearranged to give
\begin{equation}
\label{eq:NLsysRes}
\hat\bu' =\left(-i\omega  -  \left.\frac{\partial \bg}{\partial \bu}\right|_{\bu_0}\right)^{-1} \hat \bbf \defeq \mathcal{H}\hat\bbf,
\end{equation}
% Here the resolvent operator is defined to be
% \begin{equation}
%     \mathcal{H} = \left(-i\omega  -  \left.\frac{\partial \bg}{\partial \bu}\right|_{\bu_0}\right)^{-1},
% \end{equation}
with $\mathcal{H}$ being the resolvent operator associated with this equation, defined
for all $\omega$ for which the inverse in equation \ref{eq:NLsysRes} exists. 
We now describe the specific forms of this operator for parallel shear flows, starting with the incompressible case. The parallel flow assumption means that we have homogeneity in both the streamwise ($x$) and spanwise ($w$) directions (with $y$ being the wall-normal direction). Accordingly, we take spatial Fourier transforms in these directions (with corresponding wavenumbers $k_x$ and $k_z$) as well as in time.

For incompressible parallel shear flow, we express the resolvent formulation in wall-normal velocity ($v$) and vorticity ($\eta$) coordinates. Here, the incompressible Navier--Stokes equations linearized about a mean state $U(y)$ can be written as 
\begin{equation}
\label{eq:ResolventNSE1}
\begin{pmatrix}
\hat v \\
\hat \eta \end{pmatrix} = 
\underbrace{\begin{pmatrix}
-i\omega  + \Laplace^{-1} \mathcal{L}_{OS} & 0 \\
i k_z U_y &-i\omega  + \mathcal{L}_{SQ}
\end{pmatrix}^{-1}}_{\mathcal{H}_{v\eta}}
\begin{pmatrix}
\hat f_v \\
\hat f_\eta \end{pmatrix}.
\end{equation}
where the Fourier-transformed wall-normal vorticity $\hat \eta = i k_z \hat u - i k_x \hat w$, with $u$ and $w$ denoting the velocity components in the streamwise and spanwise directions, respectively. Equation \ref{eq:ResolventNSE1} is written in terms of the Orr-Sommerfeld and Squire operators, defined by
\begin{align}
\mathcal{L}_{OS} &= i k_x U \Laplace - i k_x U_{yy} - \frac{1}{Re} \Laplace^2, \label{eq:os} \\
\mathcal{L}_{SQ} &= ik_x U - \frac{1}{\Rey}\Laplace. \label{eq:sq}
\end{align}
Here the subscript $y$ denotes a derivative in the wall-normal direction, and $\Laplace = \mathcal{D}_yy -k_x^2-k_z^2$ is the Laplacian operator. 
It will also be instructive to consider an alternative expression for the resolvent operator, obtained by expressing the matrix inverse in terms of the inverses of its block components \citep{rosenberg2018efficient} 
\begin{align}
\label{eq:ResolventNSE}
\mathcal{H}_{v\eta} &= \begin{pmatrix}
\mathcal{H}_{OS}  & 0 \\
i k_z \mathcal{H}_{SQ} U_y \mathcal{H}_{OS}  & \mathcal{H}_{SQ}
\end{pmatrix}, \\
% \begin{pmatrix}
% \hat v \\
% \hat \eta \end{pmatrix} &= 
% \underbrace{\begin{pmatrix}
% \mathcal{H}_{OS}  & 0 \\
% i k_z \mathcal{H}_{SQ} U_y \mathcal{H}_{OS}  & \mathcal{H}_{SQ}
% \end{pmatrix} }_{\mathcal{H}}
% \begin{pmatrix}
% \hat f_v \\
% \hat f_\eta \end{pmatrix} \\
%
\mathcal{H}_{OS} &=  \left(-i\omega + \Laplace^{-1}\left[i k_x U \Laplace - i k_x U_{yy} - \frac{1}{Re} \Laplace^2\right]\right)^{-1} ,\label{eq:Hos} \\
\mathcal{H}_{SQ} &= \left(-i\omega + ik_x U - \frac{1}{\Rey}\Laplace\right)^{-1}.
\end{align}
where $\mathcal{H}$ is the resolvent operator, expressed in terms of the resolvent Orr-Sommerfeld ($\mathcal{H}_{OS}$) and Squire ($\mathcal{H}_{SQ}$) components. 
Here, the incompressible turbulent mean profiles are obtained by assuming an eddy viscosity model \citep{reynolds1967stability}. 
To instead express the state and forcing in terms of primitive variables (velocity components), we can apply the transformations
\begin{align}
\begin{pmatrix}
\hat u \\
\hat v \\
\hat w 
\end{pmatrix}
&= \underbrace{\mathcal{C H}_{v\eta}\mathcal{B}}_{\mathcal{H}}
\begin{pmatrix}
\hat f_u \\
\hat f_v \\
\hat f_w 
\end{pmatrix}
% \underbrace{k^{-2}\begin{pmatrix}
%     k_x\sD_y & -i k_z \\
%     k^{2} & 0 \\
%     i k_z \sD_y& i k_z
% \end{pmatrix}}_{\mathcal{C}}
% \begin{pmatrix}
% \mathcal{H}_{OS}  & 0 \\
% i k_z \mathcal{H}_{SQ} U_y \mathcal{H}_{OS}  & \mathcal{H}_{SQ}
% \end{pmatrix}
% \underbrace{\begin{pmatrix}
%     -i k_x \Laplace^{-1} \sD_y  & -k^2 \Laplace^{-1} & -ik_z \Laplace^{-1} \sD_y \\
%     i k_z & 0 &-i k_x
% \end{pmatrix}}_{\mathcal{B}},
\end{align}
where
\begin{equation}
  \mathcal{B} =   \begin{pmatrix}
    -i k_x \Laplace^{-1} \sD_y  & -(k_x^2+k_z^2) \Laplace^{-1} & -ik_z \Laplace^{-1} \sD_y \\
    i k_z & 0 &-i k_x
\end{pmatrix}, \ \
\mathcal{C} = \begin{pmatrix}
    k_x\sD_y & -i k_z \\
    (k_x^2+k_z^2)  & 0 \\
    i k_z \sD_y& i k_z
\end{pmatrix}.
\end{equation}
%Expressing the state in terms of velocity components allows for  standard norm

To explore the broader applicability of our proposed method, we will also consider the equivalent formulation for the compressible Navier--Stokes equations, though for brevity we delay a description of the equivalent compressible operator to Section \ref{sec:ResultsComp} and Appendix 1. 

The resolvent methodology proceeds by considering a singular value decomposition (SVD) 
\begin{equation}
    \mathcal{H}= \sum_{j = 1}^\infty \sigma_j \psi_j\phi_j^*, 
\end{equation}
where $\cdot^*$ denotes the adjoint, and the terms in the sum are ordered such that $\sigma_j  {\geq} \sigma_{j+1}$ for all $j$. The leading left ($\psi_ {1}$) and right ($\phi_1$) singular vectors giving the resolvent response and forcing modes corresponding to largest energy amplification (quantified by the leading singular value, $\sigma_1$). The focus of this work will be to 
discover which blocks of the resolvent operator, as formulated in equation \ref{eq:ResolventNSE} for the incompressible case, are primarily responsible for the emergence of this leading mode, describing the dominant linear amplification mechanism. The numerical discretization of the resolvent operator is obtained using a Chebyshev collocation method, utilizing the Matlab toolbox of \cite{weideman2000matlab} to form the associated differential operators. All discretizations of the resolvent operator use at least 64% change to 49
collocation points, which was found to be sufficient to give converged results.

\subsection{Block-sparsification of the resolvent operator}
\label{sec:resSparse}
We now apply this general approach described in Section \ref{sec:sparse} to the resolvent analysis framework developed in \ref{sec:resolvent}. 
Starting with the incompressible case, if we introduce coefficients $c_j$ within each sub-block of equation \ref{eq:ResolventNSE}, we can seek a reduction of this equation by finding a simplified approximate operator $\mathcal{H}_{a,v\eta}$ that minimizes the cost function
\begin{equation}
\label{eq:J}
\mathcal{J}({\bc}) =  \left\| \mathcal{C}\left( \mathcal{H}_{v\eta}- \mathcal{H}_{a,v\eta}({ \bc } )\right)\mathcal{B}\right\|_2 + \lambda \sigma_1 \left\|{ \bc } \right\|_1%\\
%& = 
\end{equation}
where $\|\cdot \|_2$ refers to the operator norm, $\bc = (c_{11}, c_{21},c_{22})^T$, and $\mathcal{H}_{a,v\eta}$ is given by  
\begin{equation}
\mathcal{H}_{a,v\eta} =  \begin{pmatrix}
{ c_{11} }\mathcal{H}_{OS}  & 0 \\
%\vspace{-0.1cm}
{ c_{21} }i k_z \mathcal{H}_{SQ} U_y \mathcal{H}_{OS}  & {c_{22} }\mathcal{H}_{SQ}
\end{pmatrix}
\end{equation} 
Note that the operator 2-norm has a direct connection with the leading singular value, as $\|\mathcal{C}\mathcal{H}_{v\eta}\mathcal{B}\|_2 = \sigma_1(\mathcal{CH}_{v\eta}\mathcal{B})$. The inclusion of the $\mathcal{B}$ and $\mathcal{C}$ operators means that the norm is directly related to kinetic energy. We choose to perform this optimization in wall normal velocity and vorticity coordinates since this naturally already allows for a compressed representation of the dynamics. % could elaborate cv primative formulation
% edit this:
As was the case in equation \ref{eq:opt2}, the first term on the right-hand side of equation \ref{eq:J} represents a measure of the difference between the original and sparsified equations, and the second term penalizes the $1$-norm of the coefficient vector $  \bc$, which promotes a solution where some components of $\bc$ are zero. %{The parameter $\lambda$ controls the tradeoff between the sparsity of $\bc$ and accuracy of the approximation $\mathcal{H}_a$, with a larger $\lambda$ giving a more sparse approximation.}
The inclusion of $\sigma_1$ in the $1$-norm term in equation \ref{eq:J} is a normalization to ensure that $\lambda$ can be set within a similar range across all spatio-temporal scales and 
Reynolds numbers.

As before, once the nonzero elements are identified, we may perform an updated optimization to fine tune the values of these nonzero coefficients such that they minimize
\begin{equation}
\label{eq:J2}
\mathcal{J}_u({\bc}) =  \left\| \mathcal{C}\left( \mathcal{H}_{v\eta}- \mathcal{H}_{a,v\eta}({ \bc } )\right)\mathcal{B}\right\|_2. %\\
%& = 
\end{equation}
A similar formulation of this sparsity-promoting optimization procedure can be formulated for the compressible case. However, as the resolvent operator is expressed in terms of three velocity components and two thermodynamic variables in this case, we instead end up with $5\times5=25$ total blocks that can each have their corresponding coefficient adjusted in the optimization procedure.
 {For the incompressible case, We emphasize that while the operator $\mathcal{CH}_{v\eta}\mathcal{B}$ maps between the three components of velocity, we still introduce sparsification in velocity-vorticity coordinates within the $\mathcal{H}_{v\eta}$ operator. Compared to considering the blocks of $\mathcal{CH}_{v\eta}\mathcal{B}$ directly, this reduces the number of coefficients required in $\bc$ from $3\times3=9$ to 3.}
    
\section{Results}
\label{sec:results}
\subsection{Sparsification of a discretized differential operator}
\label{sec:resultsD}
Before applying our proposed sparsity-promoting methodology to study fluid flows, we first provide an example where it can be applied upon a simple matrix with known properties. For this purpose, we will consider matrices that approximate the differential operator $\sD$, where
\begin{equation}
    \sD g(x) = \frac{d g}{dx}.
\end{equation}
 We will consider a discretized approximation of $\sD$ by a matrix $\mD$, where the once-differentiable function $g(x)$ is approximated by a vector $\bg$. We define this function over a periodic domain $x\in[0,2\pi)$, with the discretization using $N$ uniformly spaced points within this domain (with grid spacing $\Delta x = 2\pi/N$). There are a number of discretization schemes that can be used to define $\mD$. To start with, we consider a spectral differentiation matrix, which exactly captures the derivative assuming Fourier interpolation of the function and derivative. This matrix is dense, with entries given by
\begin{equation}
\mD_{ij} = 
\begin{cases}
\frac{1}{2} (-1)^{i-j} \cot\left(\frac{\pi (i-j)}{N}\right), & i \neq j, \\
0, & i = j,
\end{cases}
\end{equation}
for cases where $N$ is even. See \cite{Trefethen:2000} for the derivation and further discussion of the properties of this operator. Differentiation on a finite set of points can also be approximated using finite difference methods. While not possessing spectral convergence properties, such methods produce sparse differentiation matrices that are banded about the main diagonal. For example, a three-point central-differencing method gives the differentiation operator
\begin{equation}
    \mD_{ij} =
\begin{cases}
\frac{1}{2\Delta x}, & j = i+1 \ (\text{mod } N), \\
-\frac{1}{2\Delta x}, & j = i-1 \ (\text{mod } N), \\
0, & \text{otherwise}.
\end{cases}
\end{equation}
Therefore, a useful test of the general methodology described in Section \ref{sec:sparse} is to assess whether it can obtain reasonable sparse approximations to dense differentiation operators. To proceed, we introduce two slight variants of the optimization problem formulated in equation \ref{eq:opt2}. The first is to consider the optimization problem 
\begin{equation}
\label{eq:opt3}
    \mathcal{J}(\mD_a) = \|\mD \mQ - \mD_a \mQ\|_F + \lambda \|\text{vec}(\mD_a)\|_1.
\end{equation}
Here the subscript $F$ denotes the Frobenius norm, and the matrix 
\[\mQ = \left[ \bq_1, \ \bq_2 \cdots \bq_r\right]\]
is a set of $r \leq N$ test functions that are used to assess the accuracy of the approximating operator. Note that if $\mQ$ consists of an orthonormal set of basis functions spanning the computational domain, then the inclusion of $\mQ$ does not affect the optimization problem at all, as it simply amounts to a coordinate change. However, the optimization problem is modified if fewer than $N$ basis functions are used to assemble $\mQ$. For example, we could choose to assess the accuracy of $\mD_a$ on a set of Fourier modes with wavelength greater than a specified cutoff. 
 {While we find that the use of $\mQ$ can improve the results here, it will not be needed for the block-wise sparsificaton of the resolvent operator, since in that case each block still utilizes the same numerical differention schemes.} 
The second variant involves taking the solution to the optimization problem in equation \ref{eq:opt3}, and performing an additional optimization without the 1-norm term, over the identified nonzero entries of $\mD_a$ (analogous to equation \ref{eq:J2} for analysis of the resolvent operator). That is, we minimize
\begin{equation}
\label{eq:optUpdate}
     \mathcal{J}_u\m(\mD_{s,u}) = \|\mD \mQ_u - \mD_{s,u} \mQ_u\|_F
\end{equation}
subject to the constraint that the updated sparse matrix $\mD_{s,u}$ has the same sparsity pattern as $\mD_{s}$.
 Here the choice of test functions contained in the columns of $\mQ_u$ may be different to those in $\mQ$.

\begin{figure}
         % from Example_fourdif_cvx_mod.m
   \centering  \includegraphics[width=0.95\textwidth]{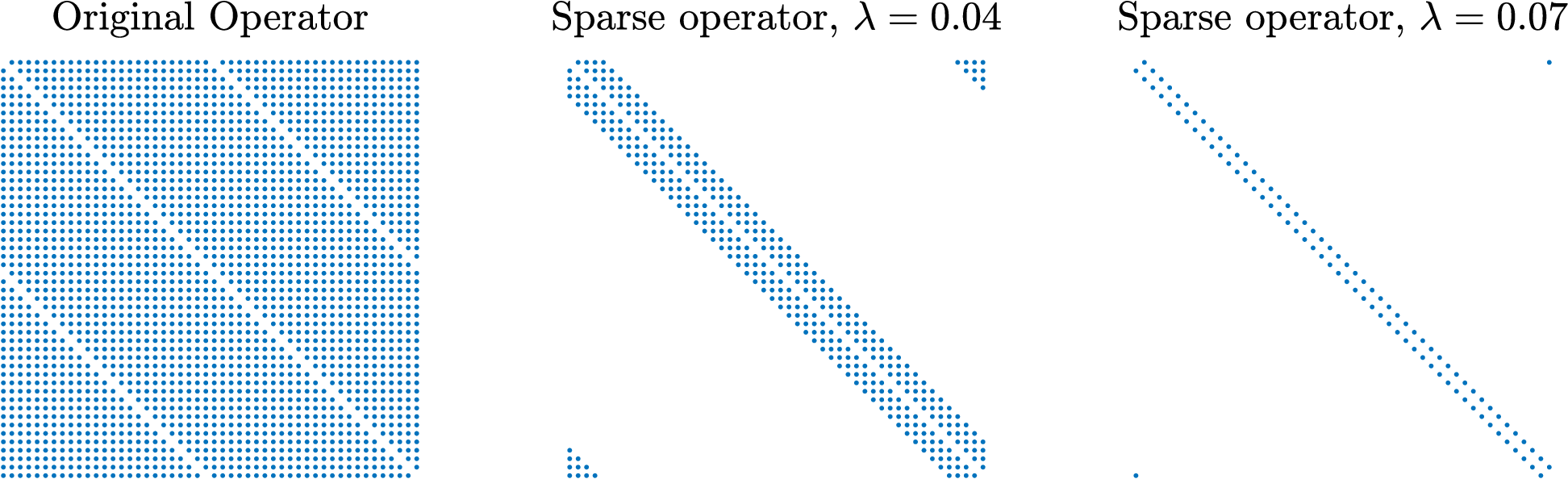}
        \caption{Structure of the spectral matrix representation of a first derivative operator (left) and identified sparse approximations of this operator obtained by optimizing equation \ref{eq:opt2} with sparsity parameters $\lambda = 0.04$ and $0.07$.
    %    sparse matrices identified when approximating a dense matrix differential operator
         }
    \label{fig:sparseD}
    \end{figure}

\begin{figure}
 % from Example_fourdif_cvx_mod.m
   \centering  (a)\includegraphics[width=0.45\textwidth]{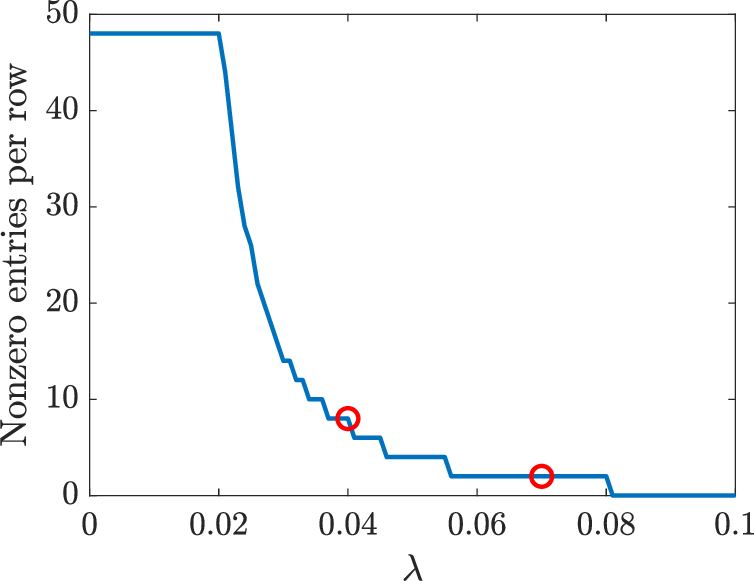}
(b)\includegraphics[width=0.45\textwidth]{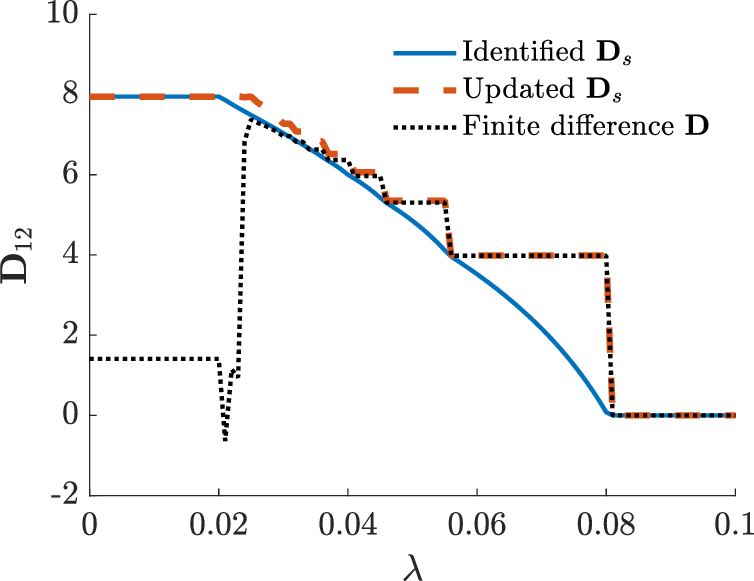}
        \caption{(a) Sparsity of identified operator as a function of the sparsity tuning parameter $\lambda$. Red circles correspond to the values of $\lambda$ shown in figure \ref{fig:sparseD}. (b) Values of the $(1,2)$ entry of the identified sparse operator both before and after applying the update step (equation \ref{eq:opt3}), in comparison to the equivalent entries in a standard finite difference scheme with the same sparsity pattern.
         }
    \label{fig:sparseD2}
    \end{figure}

We show in figure \ref{fig:sparseD} the results of solving this optimization, in terms of the structure of the identified sparse differentiation operators, using a discretization with $N=50$ points, and letting $\mQ = \mI$ in equation \ref{eq:opt3}. 
The observed sparse matrix structure matches that for finite difference stencils, featuring nonzero entries that are banded about the mean diagonal, with the number of nonzero entries dependent on the sparsity parameter. This is further shown in figure \ref{fig:sparseD2}(a), where we show the number of nonzero entries in each row of the matrix, as a function of the sparsity parameter, $\lambda$. For sufficiently small $\lambda$, the sparsity pattern is unchanged from the original operator (with 48/50 nonzero entries per row). As $\lambda$ increases, the identified operator becomes more sparse, with all entries eventually set to zero for $\lambda > 0.08$. All intermediate values of $\lambda$ result in the same banded structure observed in the cases shown in figure \ref{fig:sparseD}. To check whether the values of the nonzero entries of $\mD_a$ are also consistent with known schemes, figure \ref{fig:sparseD2}(b) shows the value of the (1,2) entry of the identified sparse matrix, both before and after performing the updated optimization in equation \ref{eq:optUpdate}. For this update step, we choose the update test functions in $\mQ_u$ to be the first $r$ Fourier modes, where $r$ is set to be the number of nonzero entries per row in the identified sparse operator. 
Also shown in figure \ref{fig:sparseD2}(b) is the equivalent entry from a matrix applying a standard finite differencing approach with the same sparsity pattern as the sparse operator identified for a given $\lambda$. For sparse operators with fewer than 10 nonzero entries per row, we observe that the update step identifies a sparse operator that closely matches the finite difference operator. This confirms that the optimization procedure is able to rediscover finite differencing operators when tasked with finding a sparse approximation to a dense differentiation operator. Furthermore, it also shows the advantage of applying the update step when performing the optimization procedure, as otherwise the sparse approximations are both less accurate and more sensitive to the specific choice of $\lambda$. 
For smaller values of $\lambda$ for which the obtained operator is less sparse, there is greater discrepancy with the finite difference operator, though this is within a regime corresponding to high order finite difference methods, would be impractical to use, particularly for problems on periodic domains. 

Although not explored further here, it is possible that this methodology could be utilized to find optimally sparse representations of more complex differential operators, and combinations thereof. 
 As our primary interest is the application of this method to study the physics of wall-bounded turbulent shear flows, further analysis of the algorithmic properties of this methodology will be considered in future work.

\subsection{Incompressible turbulent channel flow}
  \label{sec:ResultsIncomp}
         \begin{figure}
         %quickExamplePaper.m
   \centering        \includegraphics[width=0.6\textwidth]{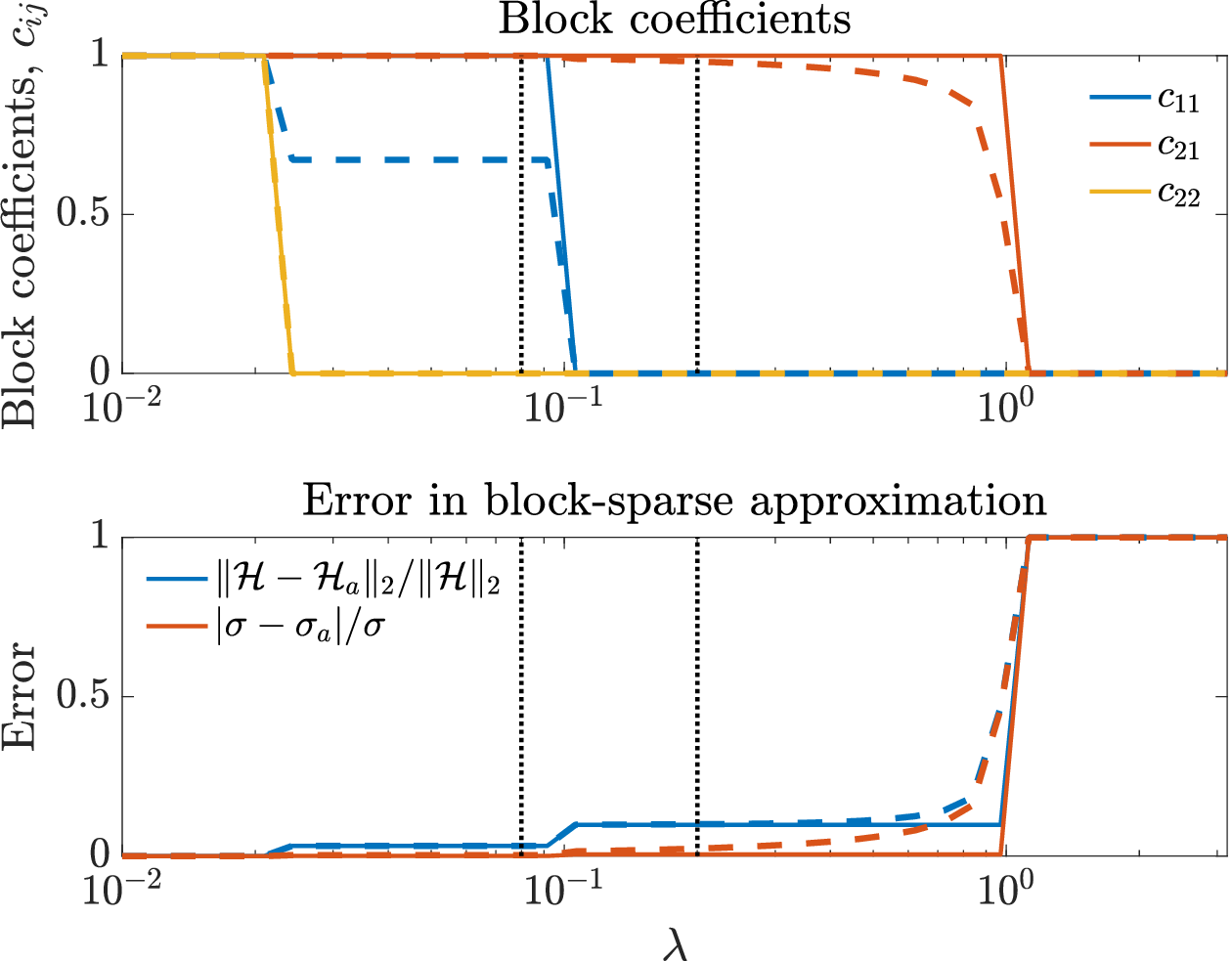}
        \caption{Results obtained from optimizing equation \ref{eq:J} (dashed lines) and after applying the coefficient update by optimizing equation \ref{eq:J2} (solid lines) for various values of the sparsification parameter, $\lambda$. 
         Results are for incompressible turbulent channel flow at a friction Reynolds number $Re_\tau = 1000$, streamwise and spanwise wavenumbers $k_x = \frac{\pi}{6}$ and $k_z = \frac{2\pi}{3}$, and a wavespeed (in inner units) $c^+ = \omega/k_x = 20$. Dotted vertical lines indicate $\lambda$ values considered in figure \ref{fig:modesincomp}
         }
    \label{fig:ex}
    \end{figure}

        \begin{figure}
        %quickExamplePaper.m
   \centering        \includegraphics[width=0.7\textwidth]{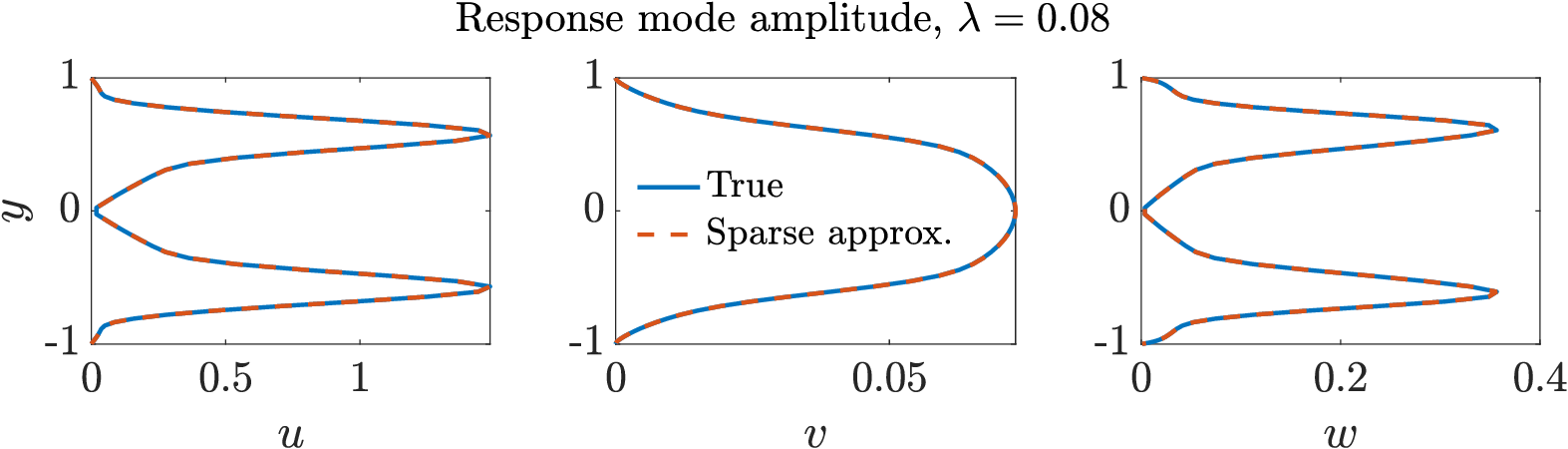} 
   \includegraphics[width=0.7\textwidth]{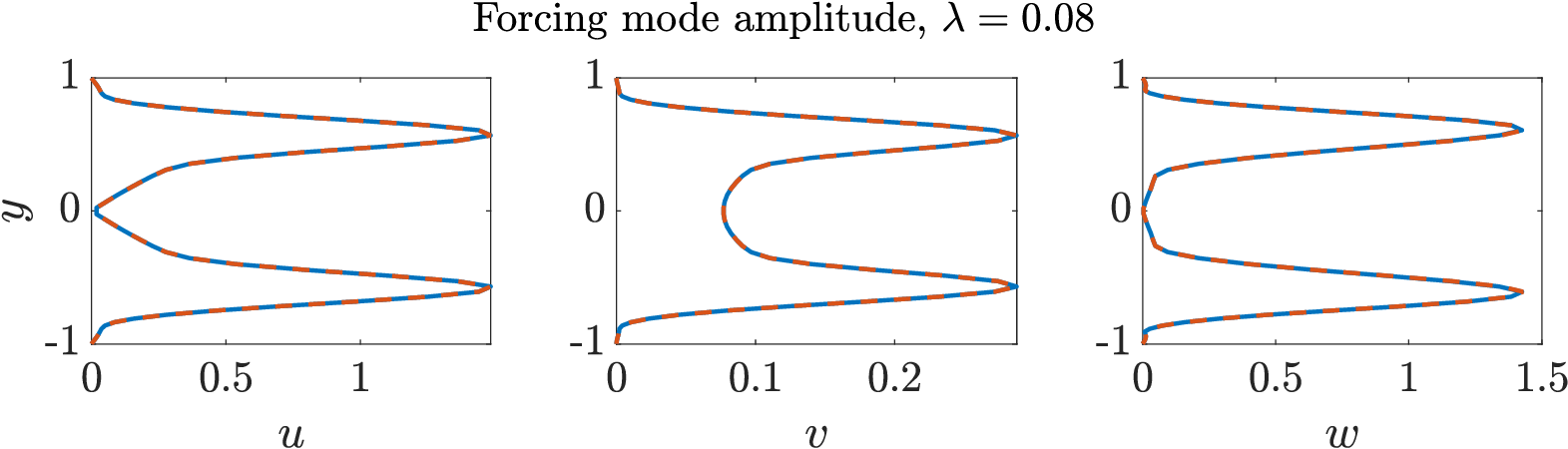} 
   \includegraphics[width=0.7\textwidth]{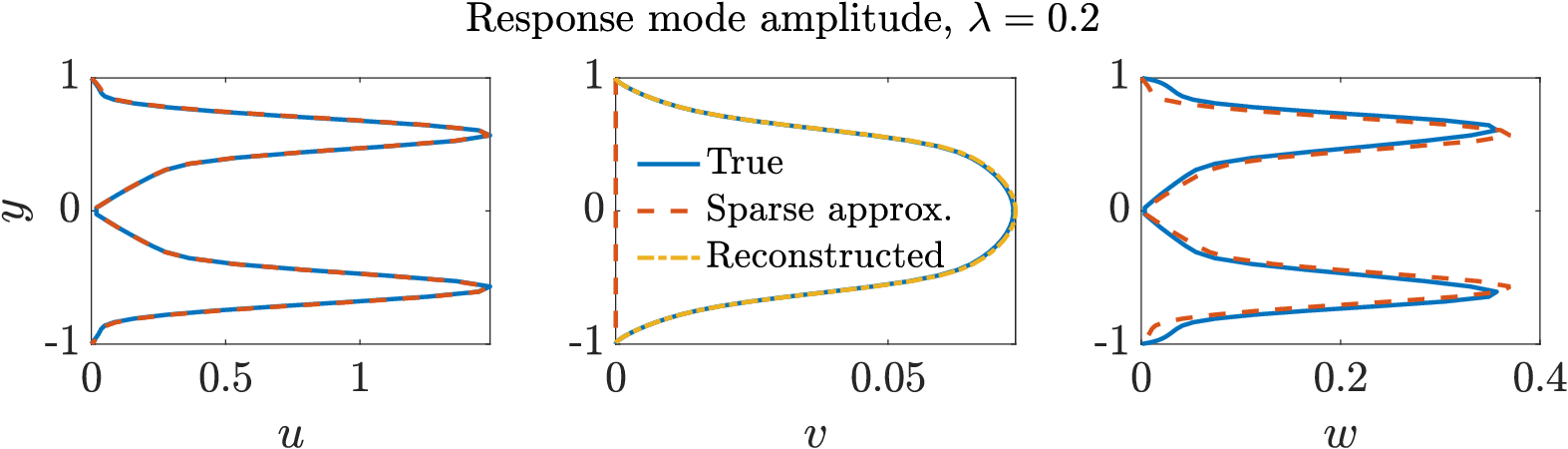}
     \includegraphics[width=0.7\textwidth]{ModesForcingLambda1.eps} 
        \caption{Leading resolvent  {forcing and} response mode shapes for both the full resolvent operator, and for sparse approximations obtained using $\lambda = 0.08$ (top) and $0.2$ (bottom). All parameters are the same as those used for figure \ref{fig:ex}.
         }
    \label{fig:modesincomp}
    \end{figure}

We now apply the sparsity-promoting methodology to resolvent operators corresponding to incompressible turbulent channel flow, following the methodology described in Section \ref{sec:resSparse}. To begin with, we
focus on structures at a specified set of spatio-temporal scales at a friction Reynolds number $Re_\tau = 1000$, where the chosen scales roughly corresponding to the largest coherent structures expected to arise in such flows.  {Here $Re_\tau = \frac{hu_\tau}{\nu}$, where $h$ is the channel half height, $\nu$ is the kinematic viscosity, and $u_\tau = \sqrt{\overline{\tau}_w/\rho}$ is the friction velocity, where $\overline{\tau}_w$ is the mean wall shear stress and $\rho$ is the fluid density.} 
 In figure \ref{fig:ex} the results of optimizing equation \ref{eq:J} are shown as a function of the sparsification parameter, $\lambda$. 
Shown are the identified block coefficients, as well as the error in the approximation. We consider two forms of error, the relative error in the estimation of the leading singular value, and the relative error between the true and approximate operators, equivalent to
 \begin{equation}
\epsilon = \sigma_1^{-1}\|\mathcal{H}-\mathcal{H}_a\|_2 = \sigma_1^{-1}\left\|
\mathcal{C}\left( \mathcal{H}_{v\eta}- \mathcal{H}_{a,v\eta}\right)\mathcal{B}\right\|_2 %\frac{\|\mathca\mathcal{H}-\mathcal{H}_a\|_2}{\|\mathcal{H}\|_2}.
\end{equation} 
For very small and large $\lambda$ the approximate operator comes out to be either the original operator or $\bm{0}$,  {with the curves corresponding to the specific suboperators in figure \ref{fig:ex} coinciding for both small and large $\lambda$}. However, there is a region for intermediate $\lambda$ where one or two of the $c_{ij}$'s have been set to zero, but where the approximate operator possesses similar properties to the full system. As was the case in Section \ref{sec:resultsD}, performing an additional optimization (equation \ref{eq:J2}) over the space of nonzero coefficients further reduces the error, and makes the results less sensitive to the specific choice of $\lambda$.

Figure \ref{fig:modesincomp} shows a comparison between the amplitude of the velocity components of the leading resolvent  {forcing and} response modes for the true and sparsified operators, for two choices of $\lambda$, plotted over the wall-normal extent of the domain. For the larger $\lambda$, only a single non-zero block is retained.  In both cases, we observe that the sparsified operators accurately capture  {all velocity components of the forcing,} and the streamwise ($u$) and spanwise ($w$) components of the response, which both relate to the wall-normal vorticity $\hat\eta = i k_z \hat u -i k_x \hat w $. 
 The terms that are truncated are consistent with previous studies, where it has been established that for three-dimensional disturbances of large streamwise extent, the dominant linear amplification mechanism arises due to the off-diagonal $c_{21}$ term %for three-dimensional disturbances of large streamwise extent 
\citep{jovanovic2005componentwise,illingworth2020streamwise,jovanovic2021bypass}. 
   This can be explained intuitively by considering the form of equation \ref{eq:ResolventNSE}, where the off-diagonal (2,1) block features a composition of two operators. This gives two opportunities for amplification: through $\mathcal{H}_{OS}$ which maps forcing in $\hat f_v$ to a response  $\hat v$, and through $\mathcal{H}_{SQ}$ which here maps this output of $\mathcal{H}_{OS}$  to a response in $\hat \eta$. 
   
 For the $\lambda = 0.2$ case in figure \ref{fig:modesincomp} where only the (2,1) block is nonzero, the $v$-component of the response modes must be zero, as the wall-normal vorticity $\eta$ only contributes towards $u$ and $w$. However, from the preceding discussion there is a response in $\hat v$ that is an intermediate step between the forcing in $\hat f_v$ and response $\hat\eta$, which can thus be reconstructed from these forcing and response components alone. Therefore, we also plot for the   $\lambda = 0.2$ case in \ref{fig:modesincomp} the reconstructed $\hat v$ response, obtained by computing
 \begin{equation}
\psi_{1,v} = \sigma_1^{-1}\mathcal{H}_{OS} \phi_{1,v},
 \end{equation}
 where $\sigma_1$ and $\phi_{1,v}$ come from the SVD of the (2,1) block.

  	\begin{figure}
	\centering
	%(a)\includegraphics[width=2in]{ Re100LowRank}
	(a)\includegraphics[width=5in,trim={5cm 1.2cm 3.5cm 0.6cm},clip]{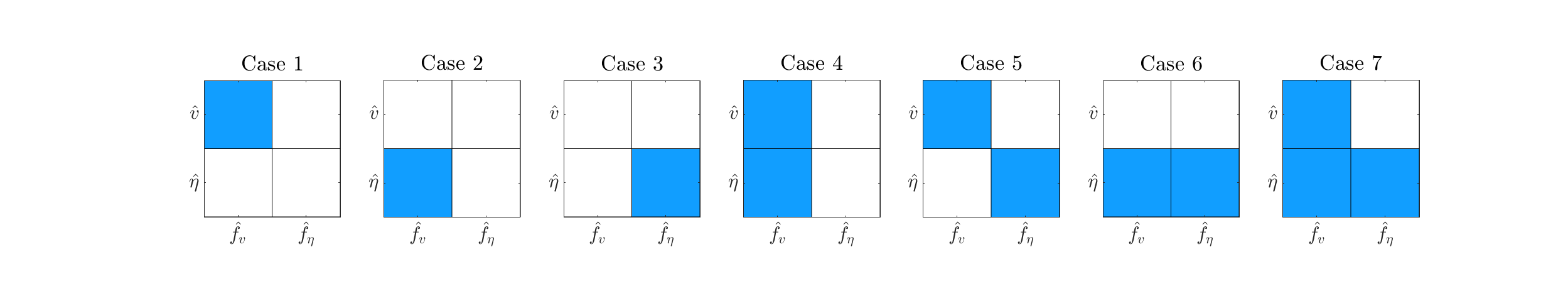}
		(b)\includegraphics[width=2.2in]{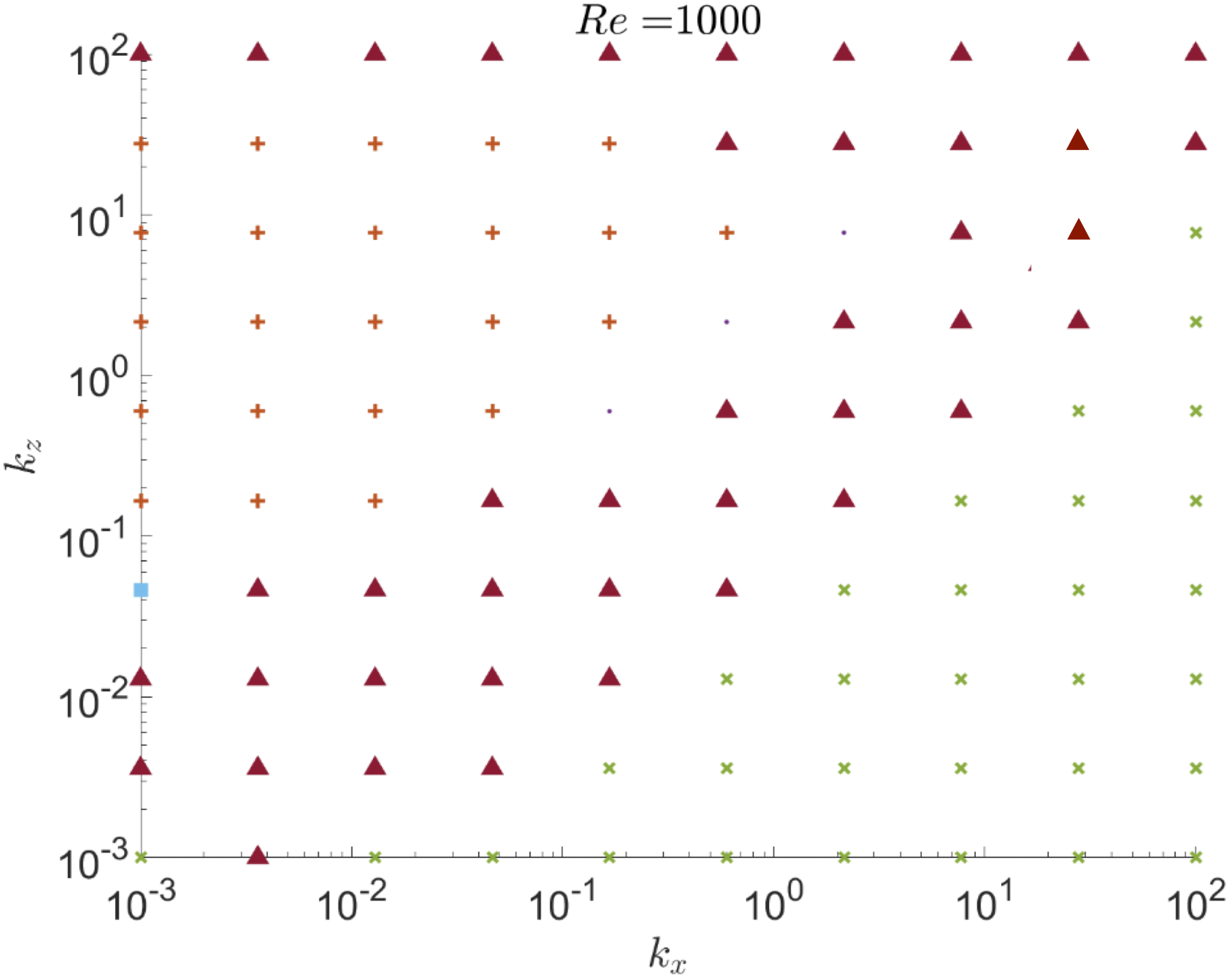}
	(c)\includegraphics[width=2.2in]{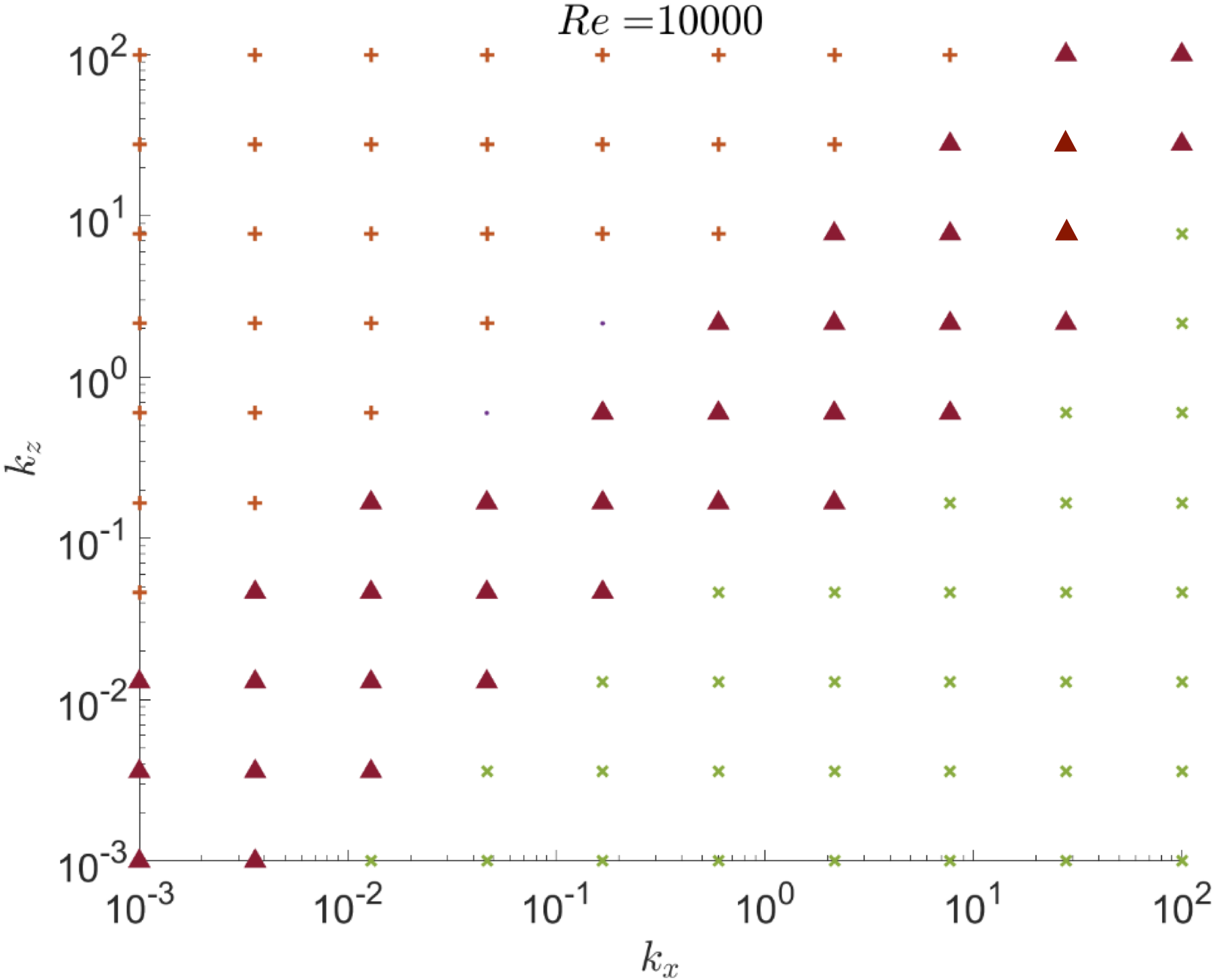}
    \includegraphics[width=0.4in]{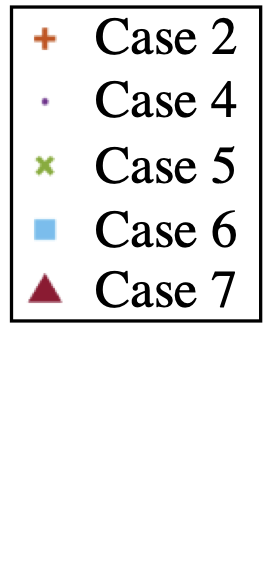}
	(d)\includegraphics[width=2.4in]{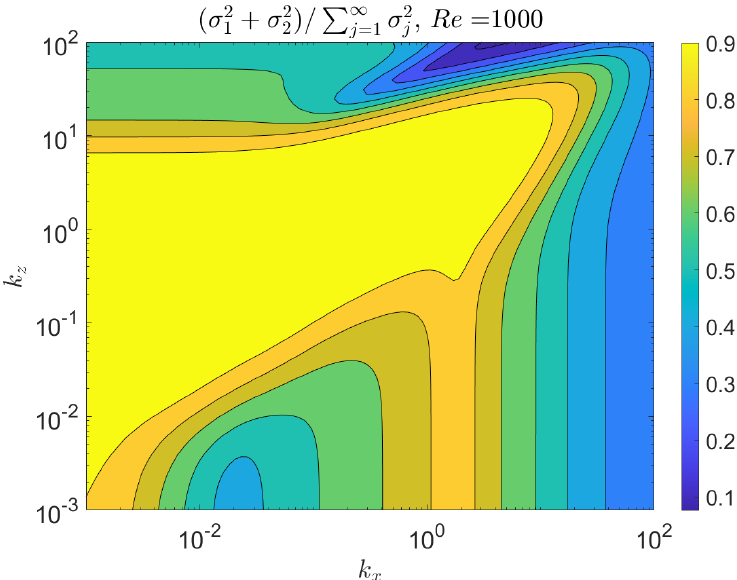}
	(e)\includegraphics[width=2.4in]{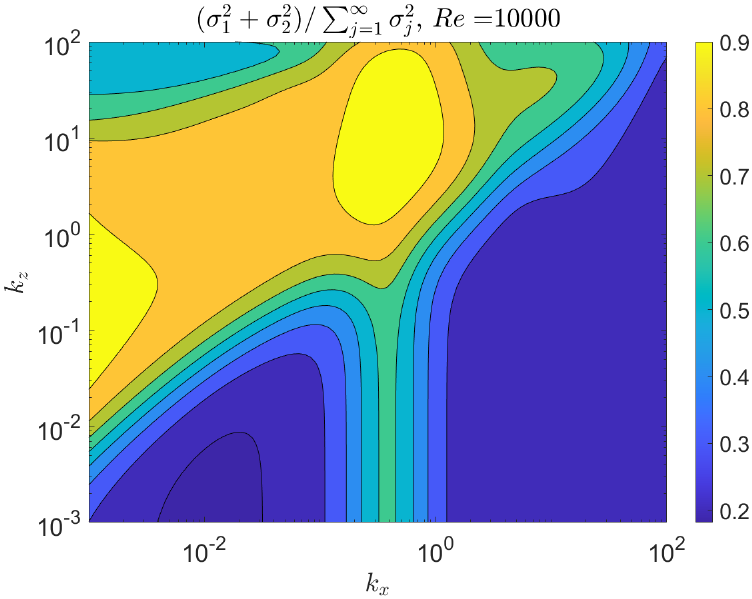}
	%(d)\includegraphics[width=2in]{ Re100Cases}
	\caption{(a) The seven possibilities for simplifying the resolvent operator in wall-normal velocity and vorticity form, where the blue entries denote retained (non-zero) blocks. Subplots  (b-c) show the results of sparsifying this operator via equation \ref{eq:J} at friction Reynolds numbers of (b) 1,000 and (c) 10,000 across a range of streamwise ($k_x$) and spanwise ($k_z$) wavenumbers. Subplots (d-e) show the extent to which the (full) resolvent operator is low rank for the parameters considered across subplots (b-c) }
	\label{fig:sweep}
	\end{figure}
 
 To test the application of this method on a wider range of scales, we now apply this sparsification method over a broader range of  streamwise ($k_x$) and spanwise ($k_z$) wavenumbers. Note that for the blockwise $2\times2$ operator considered in equation \ref{eq:ResolventNSE}, there are seven possible sparsification outcomes (excluding the trivial case where all blocks are set to zero), which are enumerated in figure \ref{fig:sweep}(a). In figure \ref{fig:sweep}(b-c), we show the sparsification that is identified for a range of $(k_x,k_z)$ pairs logarithmically spaced between $10^{-3}$ and $10^2$, at two different friction Reynolds numbers (1000 and 10,000). For simplicity, we keep the wavespeed fixed at $c^+ = \omega/k_x = 20$, which means that the critical layer (where the mean velocity is equal to this wavespeed) is also fixed. 
 For the results in figure \ref{fig:sweep}, the sparsification parameter $\lambda$ is decreased until it first produces a relative error $\epsilon < 0.1$. 
 We see that there are several distinct regions identified in both cases. In the top left of  figures \ref{fig:sweep}(b-c), we tend to identify a mechanism that was also observed in figure \ref{fig:ex}, when only the off-diagonal block of $\mathcal{H}$ is retained. Conversely, in the lower right of the figures \ref{fig:sweep}(b-c), we instead find that both diagonal blocks are retained, which corresponds to amplification via the independent effects of $\mathcal{H}_{OS}$ and $\mathcal{H}_{SQ}$. This distinction can be explained in part by the fact that the off-diagonal block is proportional to $k_z$, so is expected to be most important when $k_z \gg k_x$, and less important for amplification when $k_z \ll k_x$ In between these two regions is a diagonal band when other combinations of these blocks are selected, most notably the case where no blocks are omitted (case 7), indicating that there is no way to eliminate blocks of the original operator while still maintaining a close approximation. At the lower Reynolds number, in addition to this diagonal band, case 7 is also identified for very large $k_z$. 
 
 To give a sense of the relative importance of the different length scales considered in figure \ref{fig:sweep}(b-c), in figure \ref{fig:sweep}(d-e) we plot the proportion of the total energy captured by the leading two resolvent modes compared to the total energy across all modes at the specified wavenumbers. This quantity, which gives a measure of how close the resolvent is to being a low-rank operator, has been shown to align with the turbulent kinetic energy spectra obtained from direct numerical simulations \citep{moarref2013channels,bae2020resolvent}. 
 Equivalently, regions in these subplots where the contour levels are close to unity indicate that there is a large spectral gap in singular values after the leading two, meaning that one mechanism is dominant (we expect for singular values to often come in pairs, due to the symmetry of channel flow). 
 We observe that the low-rank region approximately coincides with the case 2 region, where only the off-diagonal block of $\mathcal{H}$ is retained in the sparsification procedure. As previously discussed, this block maps forcing in wall-normal velocity to response in wall-normal vorticity, via an intermediate response in wall-normal velocity.  This off-diagonal block is thus associated with the lift-up mechanism, where wall-normal velocity fluctuations transfer streamwise momentum towards and away from the wall.

  {Thus far, we have performed our sparsification directly upon the blocks of the resolvent operator, which corresponds to selecting a subset of input-output combinations between the wall-normal velocity and vorticity components. There are, however, other avenues through which the governing equations can be sparsified. For example, one could instead introduce coefficients directly to the $\mathcal{L}_{OS}$, $\mathcal{L}_{SQ}$ and $ik_zU_y$ terms as defined in equation \ref{eq:ResolventNSE1}, seeking an approximation of the form
 \begin{equation}
 \label{eq:alt}
 \mathcal{H}'_{v\eta,a}   =  \begin{pmatrix}
-i\omega  + c'_{11}\Laplace^{-1} \mathcal{L}_{OS} & 0 \\
 c'_{21}i k_z U_y &-i\omega  +  c'_{22}\mathcal{L}_{SQ}
\end{pmatrix}^{-1}.
 \end{equation}
 While this would perhaps make these coefficients more physically interpretable, we find that that this formulation is typically less likely to result in an accurate sparse approximation. For example, for the case considered in figure \ref{fig:ex}, we find that setting any of the coefficients $c'_{ij}$ in equation \ref{eq:alt} to zero results in a relative error close to unity. This can be understood by noting that the dominant off-diagonal term in the original formulation (equation \ref{eq:ResolventNSE1}) depends on all of the blocks of the non-inverted operator. 
 Note also that if we include the $i\omega$ terms in the multiplication by $c'_{ii}$, then setting any of these coefficients on the diagonal blocks to zero would result in a non-invertible operator.  
 Lastly, we note that this alternative formulation is potentially more difficult to optimize in practice, since the coefficients enter into the  2-norm term in the cost function in a nonlinear manner. 
 }

\subsection{Time domain analysis}
\label{sec:resultsTransient}
 {

While this work has focused on the resolvent formulation in the frequency domain, we can also consider an analogous procedure in the time domain. In particular, here we consider how the sparsification procedure can be applied to optimal transient growth analysis, which identifies the initial state that results in the largest energy amplification over a specified time horizon. 

We will again focus on the incompressible case considered in section \ref{sec:ResultsIncomp}. The time domain form of equation \ref{eq:ResolventNSE1} is given by
\begin{equation}
    \frac{\partial}{\partial t} 
\begin{pmatrix}
{v} \\
    \eta
    \end{pmatrix}
    = 
%      \begin{pmatrix}
%     \Laplace & 0 \\
%     0 & I
% \end{pmatrix}^{-1}
\begin{pmatrix}
\Laplace^{-1}\mathcal{L}_{OS} & 0 \\
i k_x U_y & \mathcal{L}_{SQ}
    \end{pmatrix}
\begin{pmatrix}
{v} \\
    \eta
    \end{pmatrix}.
    \end{equation}
 The associated finite-time propagation operator for a time horizon $\tau$ is given by
 \begin{equation}
\mathcal{F}(\tau) = \exp\left[\tau\begin{pmatrix}
\Laplace^{-1}\mathcal{L}_{OS} & 0 \\
i k_x U_y & \mathcal{L}_{SQ}
    \end{pmatrix}\right] =
\begin{pmatrix}
\exp\left[\tau\Laplace^{-1}\mathcal{L}_{OS}  \right] & 0 \\
\mathcal{F}_{21} & \exp\left[\tau\mathcal{L}_{SQ} \right]
\end{pmatrix},
 \end{equation}
 where the off-diagonal entry $\mathcal{F}_{21}$ is given by \citep{dieci2001conditioning}
 \begin{equation}
\mathcal{F}_{21} = \int_{s=0}^1
\exp\left[s\tau\mathcal{L}_{SQ} \right]
(i k_x U_y \tau)
\exp\left[(1-s)\tau\Laplace^{-1}\mathcal{L}_{OS} \right] ds.
 \end{equation}
 As for the resolvent case, we have used the block upper triangular nature of the equations to find explicit forms of the blocks of the matrix exponential. 
Analogously with resolvent analysis, the leading right and left singular vectors of $\mathcal{F}(\tau) $ give the initial and final conditions corresponding to maximal energy amplification (by a factor of $\sigma_1^2$) over the time horizon $\tau$. We seek a simplification of the form 
\begin{equation}
\mathcal{F}_a(\tau) = 
\begin{pmatrix}
c_{11}\exp\left[\tau\Laplace^{-1}\mathcal{L}_{OS}  \right] & 0 \\
c_{21}\mathcal{F}_{21} & c_{22}\exp\left[\tau\mathcal{L}_{SQ} \right],
\end{pmatrix}
\end{equation}
which minimizes the cost function (analogous to equation \ref{eq:J})
\begin{equation}
\label{eq:Jtrans}
\mathcal{J}({\bc}) =  \left\| \mathcal{C}\left( \mathcal{F}- \mathcal{F}_{a}({ \bc } )\right)\mathcal{B}\right\|_2 + \lambda \sigma_1 \left\|{ \bc } \right\|_1.%\\
%& = 
\end{equation}
As before, once the nonzero blocks are identified, we can update the values of the nonzero coefficients by minimizing
\begin{equation}
\label{eq:J2trans}
\mathcal{J}_u({\bc}) =  \left\| \mathcal{C}\left( \mathcal{F}- \mathcal{F}_{a}({ \bc } )\right)\mathcal{B}\right\|_2. %\\
%& = 
\end{equation}
Using the same parameters considered for incompressible channel flow in figure \ref{fig:ex}, figure \ref{fig:extransient} shows the results of applying the sparsification procedure to $\mathcal{F}$. Here we use a time horizon $\tau = 6.4$, which gives the largest energy amplification for this system at the specified parameters. We observe very similar results to the resolvent analysis case, where $c_{22}$ and then subsequently $c_{11}$ are set to zero as $\lambda$ increases. This suggests that the dominant mechanisms identified for energy amplification for this system are independent of whether the analysis is performed in the time or frequency domain. While not shown, we find similar results for a range of time horizons.

         \begin{figure}
         %quickExamplePaperTransientClean.m
   \centering        \includegraphics[width=0.6\textwidth]{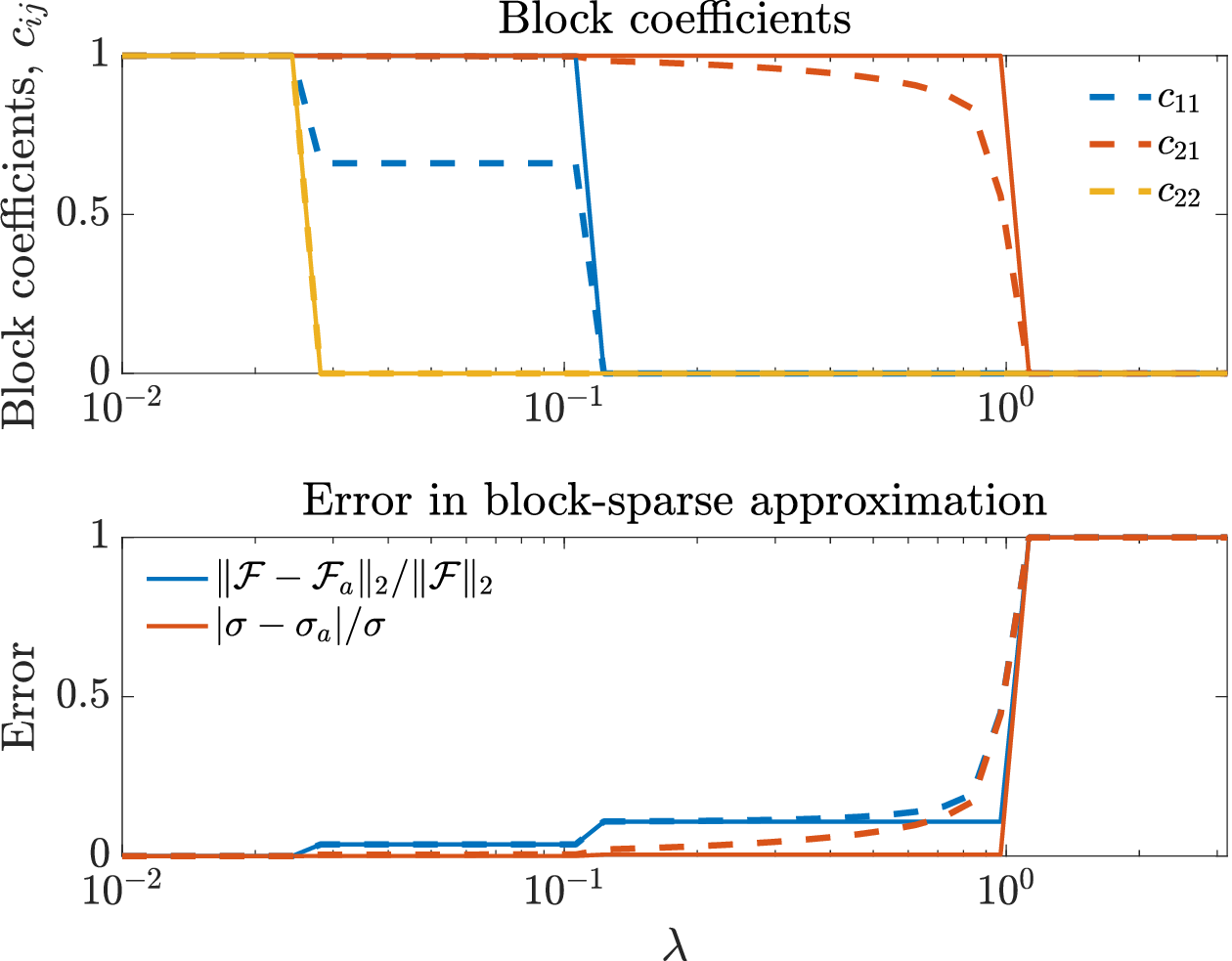}
        \caption{Results obtained from optimizing equation \ref{eq:Jtrans} (dashed lines) and after applying the coefficient update by optimizing equation \ref{eq:J2trans} (solid lines) for various values of the sparsification parameter, $\lambda$. 
         Results are for incompressible turbulent channel flow with the same parameters considered in figure \ref{fig:ex}, but for transient growth rather than resolvent analysis.
         }
    \label{fig:extransient}
    \end{figure}

}

  \subsection{Compressible Couette flow}
  \label{sec:ResultsComp}
To explore the broader applicability of the proposed methodology, we now consider laminar compressible Couette flow. 
This flow has been the subject of several previous studies that utilize a range of linear analysis methods \citep{duck1994couette,malik2006nonmodal,dawson2019couette,bhattacharjee2023structured,bhattacharjee2024structured}, making it a convenient choice for testing our methodology in the compressible regime. %
While this configuration is laminar, for linearized analyses many findings are qualitatively similar when comparing laminar and turbulent mean profiles. 
Here, the resolvent operator is formulated in terms of the velocity, density, and temperature, giving a state vector with five components, $\bq = (u,v,w,\rho,T)$. We use a compressible flow energy norm first formulated by \cite{chu1965energy}. This means that there are $5\times5$ blocks in the resolvent operator, substantially increasing the dimensionality of the optimization problems in equations \ref{eq:J} and \ref{eq:J2}. Here, the equivalent to equations \ref{eq:J} is given by that we look to minimize 
\begin{equation}
\label{eq:Jcomp}
\mathcal{J}({\bc}) =  \left\|\mathcal{H}_{comp}- \mathcal{H}_{a,comp}({ \bc) } \right\|_2 + \lambda \sigma_1 \left\|{\text{vec}(\bc) } \right\|_1%\\
%& = 
\end{equation}
where $\mathcal{H}_{comp}$ is the compressible resolvent operator, and $\mathcal{H}_{a,comp}$ the block sparse approximation, given by  
\begin{equation}
\mathcal{H}_{a,comp} =  \begin{pmatrix}
 c_{11} \mathcal{H}_{11}  & c_{12} \mathcal{H}_{12}  & c_{13} \mathcal{H}_{13}  & c_{14} \mathcal{H}_{14}  & c_{15} \mathcal{H}_{15} \\
 c_{21} \mathcal{H}_{21}  & c_{22} \mathcal{H}_{22}  & c_{23} \mathcal{H}_{23}  & c_{24} \mathcal{H}_{24}  & c_{25} \mathcal{H}_{25} \\
  c_{31} \mathcal{H}_{31}  & c_{32} \mathcal{H}_{32}  & c_{33} \mathcal{H}_{33}  & c_{34} \mathcal{H}_{34}  & c_{35} \mathcal{H}_{35} \\
   c_{41} \mathcal{H}_{41}  & c_{42} \mathcal{H}_{42}  & c_{43} \mathcal{H}_{43}  & c_{44} \mathcal{H}_{44}  & c_{45} \mathcal{H}_{45} \\
    c_{51} \mathcal{H}_{51}  & c_{52} \mathcal{H}_{52}  & c_{53} \mathcal{H}_{53}  & c_{54} \mathcal{H}_{54}  & c_{55} \mathcal{H}_{55}.
\end{pmatrix}
\end{equation} 
Here $\mathcal{H}_{ij}$ denotes the block of compressible resolvent operator that maps a forcing in the $j$-th component to a response in the $i$-th component of the state. 
As in the incompressible case, we can perform an additional optimization over the space of non-zero coefficients identified from equation \ref{eq:Jcomp}, minimizing
  \begin{equation}
\label{eq:Jcomp2}
\mathcal{J}({\bc}) =  \left\|\mathcal{H}_{comp}- \mathcal{H}_{a,comp}( \bc ) \right\|_2
%& = 
\end{equation}
the compressible resolvent operator is obtained from a linearization of the compressible Navier--Stokes equations, the components of which are given explicitly in Appendix A.
    Further details concerning the formulation of the compressible resolvent used here can be found in \cite{dawson2019couette,dawson2020prediction,bae2020resolvent}.  

        \begin{figure} % MENTION MORE VARIABLES!!
        %runSparseBlocksK10plotsPaper
   \centering
            \includegraphics[width=0.95\textwidth]{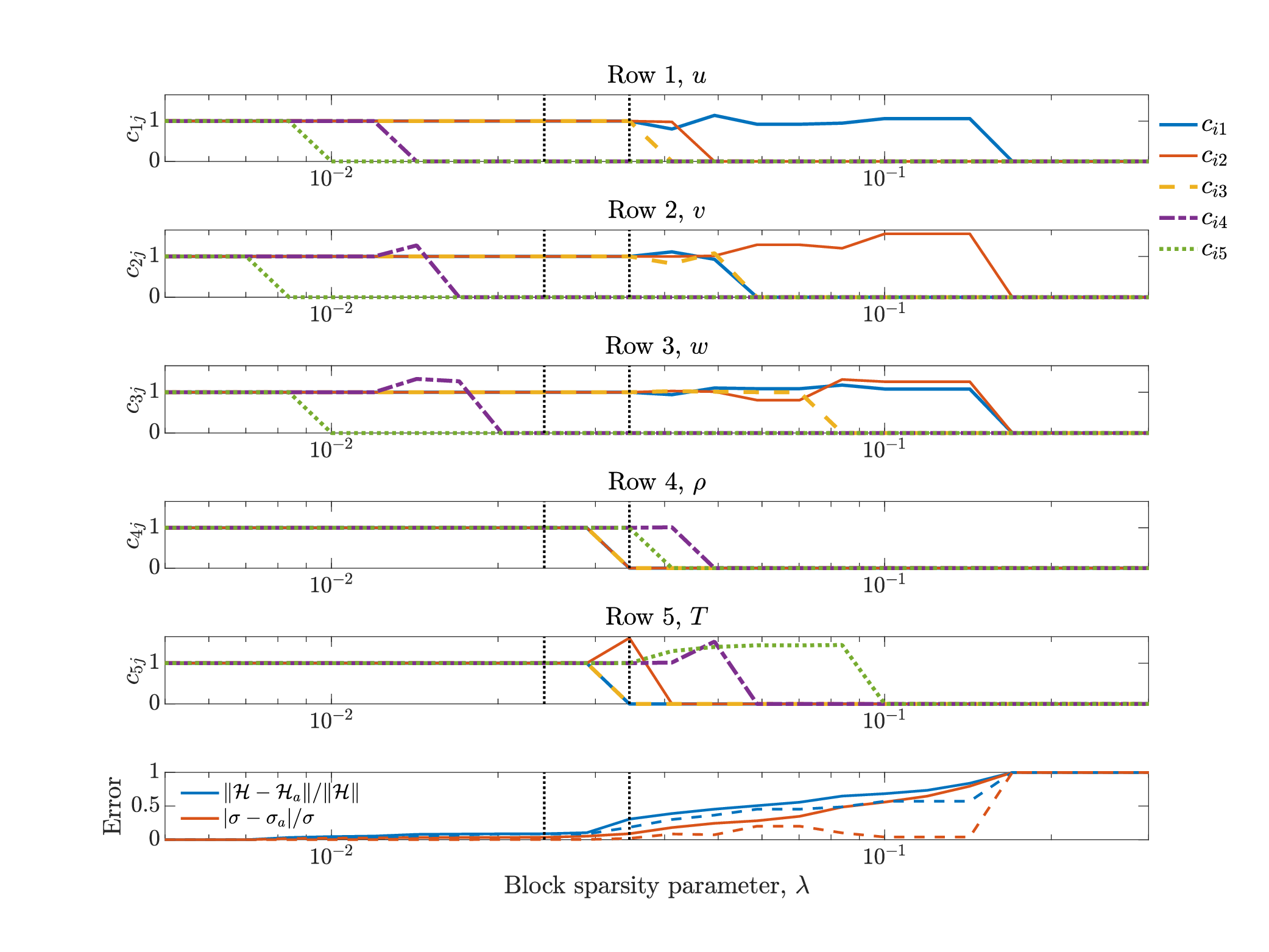}
        \caption{Results from optimizing equations \ref{eq:Jcomp} and \ref{eq:Jcomp2} for  compressible laminar Couette flow, with $Re = 1000$, $M = 2$,  $k_x = k_z = 10$, and $c = 0.5$,
        showing the elimination of several blocks coupling the dynamic and thermodynamic variables with small error in the approximate operator. The solid lines in the lowest subplot show the results of optimizing equation \ref{eq:Jcomp} alone, with the
        dashed lines showing the errors after performing the additional optimization (equation  \ref{eq:Jcomp2}).
        %\new{\ul{This demonstrates the suitability of extending the proposed methodology to the compressible regime}.}
        }
    \label{fig:exComp}
    \end{figure}

   %      \begin{figure} % MENTION MORE VARIABLES!!
   % \centering
   %          \includegraphics[width=1.0\textwidth]{ CompSparseUpdated.png}
   %      \caption{Results from optimizing equation \ref{eq:J} for  compressible laminar Couette flow, with $Re = 1000$, $M = 2$,  $k_x = k_z = 10$, and $c = 0.5$, 
   %      showing the elimination of several blocks coupling the dynamic and thermodynamic variables with small error in the approximate operator.  %\new{\ul{This demonstrates the suitability of extending the proposed methodology to the compressible regime}.}
   %      }
   %  \label{fig:exComp}
   %  \end{figure}
  
  Sample results  at a Mach number ($M$) of 2 obtained from applying the proposed sparsification methodology for a range of values of the sparsity parameter $\lambda$ are shown in figure \ref{fig:exComp}. For clarity, we only show the coefficients after performing the second optimization over the space of nonzero coefficients (equation \ref{eq:Jcomp2}).
  %This figure shows the form of the optimization problem for the compressible regime in the top left, where we now have the ability to set to zero any of the blocks corresponding to componentwise forcing and response pairs between all of the five state variables. 
  
  %The results of performing the optimization for a range of values of the sparsity parameter $\lambda$ are shown on the right, with the form of the approximate operator for $\lambda = 0.02$,  and the corresponding true and approximate leading resolvent response mode components, shown in the bottom left. 
  
  As was the case in the incompressible regime, we again find values of $\lambda$ with small approximation error, but where several of the coefficients $c_{ij}$ have been set to zero. We focus attention on two such $\lambda$ values, indicated with vertical lines in figure \ref{fig:exComp}. These cases identify sparsifying coefficients
  \begin{align}
      \bc(\lambda = 0.242) &= \begin{pmatrix}
 1 & 1 & 1 & 0 & 0 \\
 1 & 1 & 1 & 0 & 0 \\
 1 & 1 & 1 & 0 & 0 \\
 1 & 1 & 1 & 1 & 1 \\
 1 & 1 & 1 & 1 & 1
\end{pmatrix}, \\
 \bc(\lambda = 0.346) &= \begin{pmatrix}
 1 & 1 & 1 & 0 & 0 \\
 1 & 1 & 1 & 0 & 0 \\
 1 & 1 & 1 & 0 & 0 \\
 0 & 0 & 0 & 1 & 1 \\
 0 & 1.6175 & 0 & 1 & 1
\end{pmatrix}, \label{eq:lambda2comp}
  \end{align}
  with corresponding relative errors of 0.0850 and 0.1840 respectively. 
 In both cases, we observe that a number of the coupling terms between the dynamic (first three components) and thermodynamic (remaining two components) of the resolvent operator are set to zero. In particular, we find zeros in the last two columns of the first three rows of the resolvent operator in both cases. This indicates that the response in the velocity components has become independent of forcing in the density and temperature variables. This is consistent, for example, with the Morkovin hypothesis \citep{morkovin1962hypothesis}, which suggests that the dynamics of the velocity field fluctuations are largely the same as those observed in the incompressible regime. 
   This is also consistent with previous findings that the streamwise velocity component of the leading resolvent mode for compressible flow can often be accurately captured from incompressible analyses about the compressible mean field \citep{dawson2020prediction}. 
    For the larger value of $\lambda$, we obtain almost a complete decoupling between the velocity and thermodynamic variables. The retention of the wall-normal velocity component ($\mathcal{H}_{52}$) can be reasoned by considering the similarity of the dynamics between the temperature and streamwise velocity fields, as consistent with the strong Reynolds analogy \citep{morkovin1962hypothesis,smits2006supersonic}. As observed in the previous section, it is typical for large energy amplification to be associated with forcing in wall-normal velocity and response in streamwise velocity. Therefore, if the dynamics of the fluctuating temperature field are similar, then a forcing in wall-normal velocity is also expected to be important for the temperature field, as is identified here. 
    While not shown here, we observe that this decoupling between velocity and thermodynamic variables can be achieved with even smaller error as the Mach number is reduced. 
    
    Figure \ref{fig:exComp} shows certain features not present in the incompressible case, including the identification of some coefficients that are greater than 1 (as also observed in equation \ref{eq:lambda2comp}), meaning that the corresponding blocks are given a larger weighting than in the original operator. This likely indicates that these blocks are partly compensating for the effect of other dynamics eliminated in the sparsification.  {For example, for the case in equation \ref{eq:lambda2comp} when $c_{52} > 1$, figure \ref{fig:exComp} indicates that this occurs for the value of $\lambda$ where $c_{51}$ and $c_{52}$ are first set to zero. Assuming some correlation between the individual contributions of forcing in the $u$, $v$, and $w$ components on the response of $T$, increasing the size of the $c_{52}$ term could partially compensate for the elimination of $c_{51}$ and $c_{52}$. 
    }
    For larger values of $\lambda$, we also observe cases where these larger coefficients can reduce the error in the leading singular value, even when the operator error is large. 

            \begin{figure} 
        %runSparseBlocksK10plotsPaper
   \centering
        \includegraphics[width=0.8\textwidth]{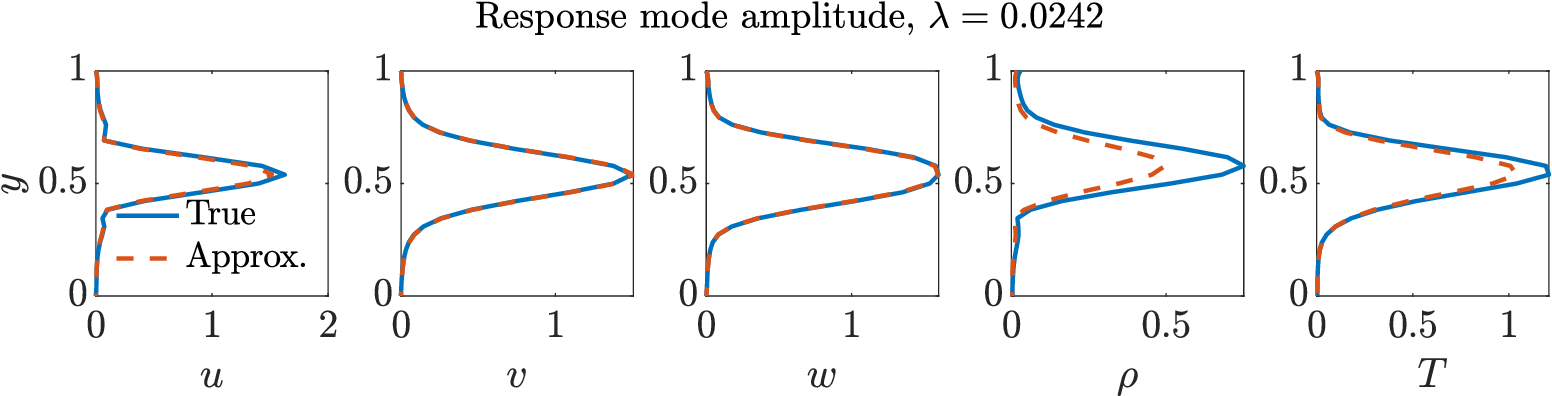} \\
        \includegraphics[width=0.8\textwidth]{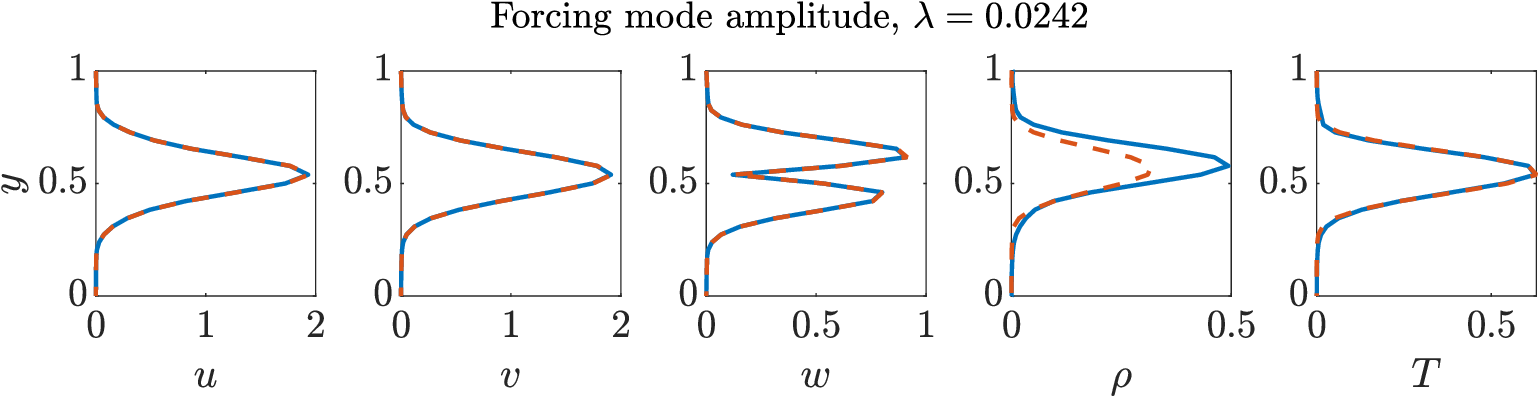} \\     \includegraphics[width=0.8\textwidth]{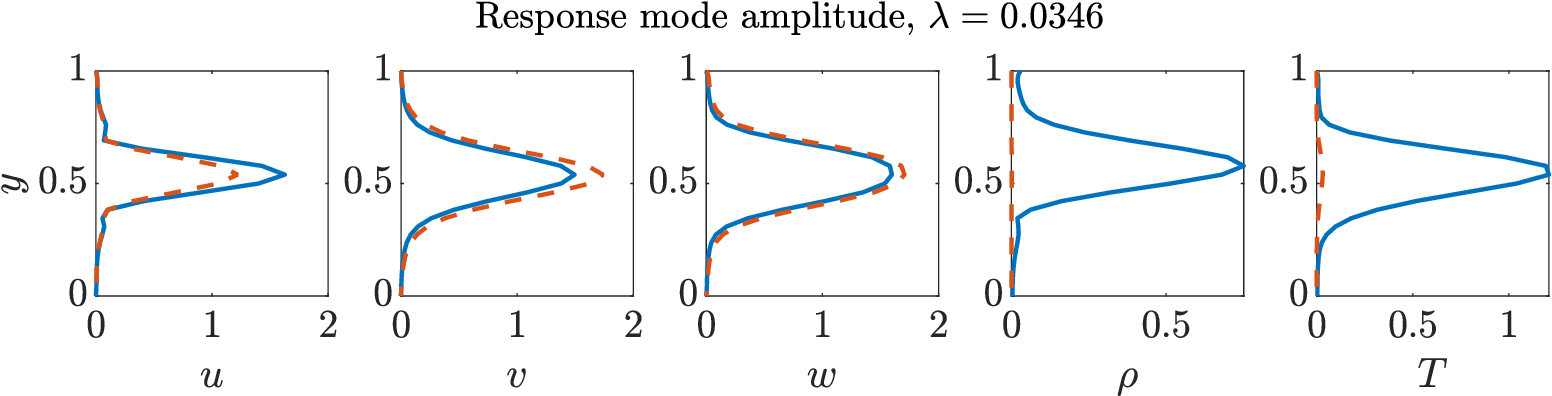}  \\
         \includegraphics[width=0.8\textwidth]{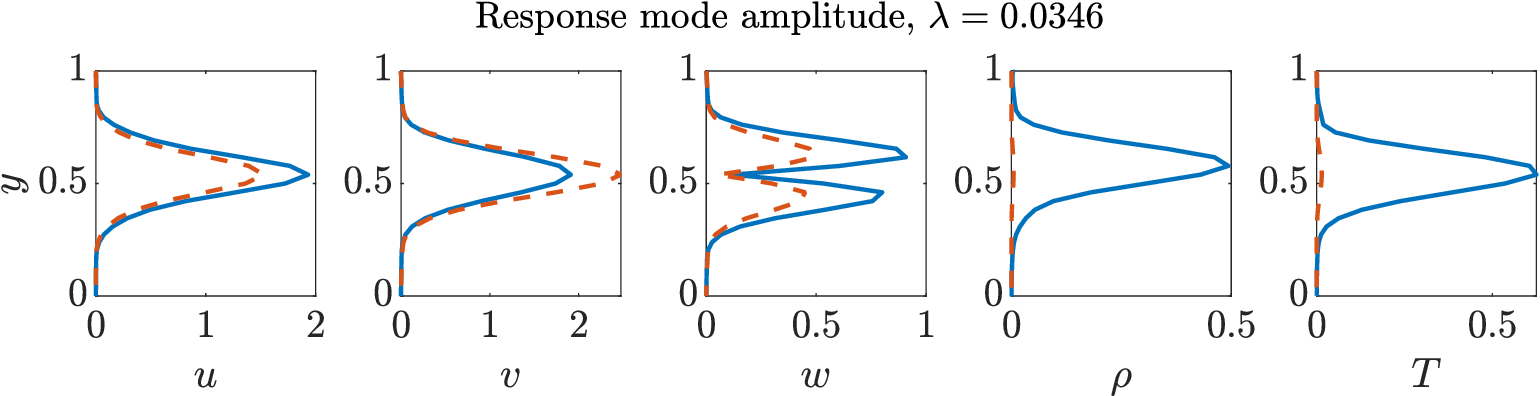}        
        \caption{Leading compressible resolvent  {forcing and} response mode components (absolute values) of the true and sparsified operators, for the values of $\lambda$ identified in figure \ref{fig:exComp}.
        }
    \label{fig:compModes}
    \end{figure}

Figure \ref{fig:compModes} shows the leading resolvent  {forcing and} response modes for the two $\lambda$ values highlighted in figure \ref{fig:exComp}. 
    For these modes, we find that the velocity components are all accurately captured by the approximate operator, particularly for the $\lambda  = 0.0242$ case. For this case, the thermodynamic variables of the approximate operator are of similar shape to those of the true operator, but with lower amplitude.  
    For $\lambda  = 0.0346$, the leading mode has velocity components that are entirely decoupled from the thermodynamic variables, which is also consistent with the prior analysis and discussion. %, indicating that some of the neglected terms have an impact on these responses.
  % Looking at the coefficients $c_{ij}$, this is likely due to several of them taking nonzero values less than one. 
   %While  not shown  here, it is possible that improved performance could be obtained by setting all nonzero coefficients to take a value of unity, indicating that the non-zero blocks are identical to those of the original operator.

 {
\subsection{Computational considerations}
\label{sec:computational}
We now discuss details pertaining to the computational methods and costs associated with minimizing the optimization problems specified by equations \ref{eq:J}, \ref{eq:Jtrans} or \ref{eq:Jcomp}, for example. As mentioned in section \ref{sec:sparse}, we utilize the SDPT3 algorithm within the Matlab CVX toolbox to minimize these objective functions. This methodology recasts these optimization problems as semidefinite programs, which are often solved using the corresponding dual formulation of the problem. %can be solved using a variety of methods.
For the resolvent analysis cases where sparsity is imposed block-wise to the operator, the dimensions of the resolvent operators is much larger than the number of optimization parameters. In such cases, the computational bottleneck involves forming and performing calculations and evaluations of the 2-norm term at each iteration of the optimization procedure,
which is similar to the computational complexity of resolvent analysis itself. Empirically, we find that the sparsity-promoting optimization for the incompressible and compressible cases takes about 10 and 80 seconds respectively on a laptop (which typically requires around 20 iterations for convergence). 
This compares to less than one second for a standard resolvent analysis. 
For the incompressible case where there are only $2^3 = 8$ combinations of sparse operators (assuming that $c_{ij}\in \{0,1\}$), it would be feasible to directly compute the cost function (or even a 0-norm variant) for all of these permutations without needing to explicitly optimize. For the compressible case however, again assuming $c_{ij}\in \{0,1\}$, there would be $2^{23} \approx 8.4\times 10^7$ such combinations (noting that there are 23 nonzero blocks in the compressible resolvent operator), making an explicit analysis of all permutations infeasible.

The formulation used in this work assumes that the resolvent operator is explicitly formed, which may not be possible for cases with two or three dimensions of spatial inhomogeneity. In such cases, it may be necessary to compute or approximate the 2-norm term without explicitly forming the resolvent operator. If it is infeasible to form and/or invert the resolvent operator, it could be possible to infer or approximate it using randomized \cite{ribeiro2020randomized} or timestepping \citep{martini2021efficient,farghadan2023scalable} methods, applied either individually to the true and approximate resolvent operators, or directly to their difference. 
Another possible alternative would be to work in the time domain following the analysis in section \ref{sec:resultsTransient}, which for the example considered here gives comparable results. 
Additionally, While the SDPT3 alogorithm was sufficient for the numerical optimization problems encountered in the present, higher-dimensional problems may necessitate the use of alteratives to interior point methods such as SDPT3. In particular, first order methods such as the alternating direction method of multipliers \citep{wen2010alternating} may be more suited in such cases. Considering such optimization problems when the state dimension is substantially larger will be the subject of future work.
}
    
     \section{Conclusions}
     \label{sec:conclusions}
     This work has introduced a novel methodology for identifying which terms within a given set of equations are the most important for retaining the properties of the original equations. This provides an objective framework for simplifying these equations through a sparsification procedure, where terms within the original equation are set to zero. This method was applied in the context of resolvent analysis, to identify simplified operators that possess similar leading singular values and vectors, corresponding to dominant linear energy amplification mechanisms. 
     
     For both incompressible and compressible wall-bounded shear flows, the method identified mechanisms that are consistent with mathematical and physical understanding of these systems. For incompressible channel flow, we find that a single block of the governing equations captures the majority of the response across a range of scales for which the resolvent operator is approximately low rank. This block is associated with forcing and response in the wall-normal velocity and vorticity components, respectively, and is associated with the lift-up mechanism. An alternative method to arrive at this conclusion could involve performing component-wise analysis of each block, as performed by \cite{jovanovic2005componentwise}. For compressible flow, we find (for one set of wavenumbers) a partial decoupling between the velocity components and thermodynamic variables,  meaning that the velocity response is largely driven by forcing in the velocity components, rather than via the thermodynamic variables.  The response in the density and temperature, however, are found to be at least partially coupled to the dynamic variables for the case considered. In particular, we find that the temperature response requires forcing in wall-normal velocity, consistent with its dynamical similarity to the streamwise velocity field via the strong Reynolds analogy. Further work could also look towards approximating the compressible equations with sparse blocks from their incompressible counterparts.
     
  In this work, the sparsification procedure was performed in a blockwise manner. Further work will look to extend this technique such that individual terms within each block can each be isolated and potentially removed. 
 It is also possible that methods designed to enable sparse signal recovery from block-sparse data \citep{eldar2010block,kramer2017sparse} could be leveraged to further analyze and extend the methodology proposed here.

 While the physical mechanisms for the flow configurations considered here are already relatively well understood, further work will apply this automated sparsification methodology to cases where the underlying physics are not as well known.   As well as obtaining physical insight, identifying simplified operators can allow for further theoretical analysis, such as the analytic prediction of leading resolvent mode shapes \citep{dawson2019shape,dawson2020prediction}. 

\section*{Acknowledgments}
This work was supported by the National Science Foundation under Award CBET-2238770.

\appendix
\section{Compressible resolvent operator}
\label{app1}
Here we show the block components of the linearized compressible Navier-Stokes equations. Assuming that the resolvent form of these equations is given by
\begin{equation}
\label{eq:rescomp}
    \hat\bq =\left(-i\omega  + \mathcal{L}_{comp}\right)^{-1} \hat \bbf,
\end{equation}
where the Fourier-transformed state $\hat\bq = (\hat u,\hat v,\hat w,\hat \rho,\hat T)$, the nonzero blockwise components of $\mathcal{L}_{comp}$ are given by
\allowdisplaybreaks
\begin{align*}
\mathcal{L}_{11} &= ik_x U + Re^{-1}T_0\left(2k_x^2\mu_0 + k_z^2 \mu_0 + k_x^2\lambda_0 -\mu_0\sD_{yy} -\mu_{0,y} \sD_y\right) \\ % changed k_z^2\lambda_0 to k_z^2\lambda_0  11/5/18
\mathcal{L}_{12} &= U_y-ik_x  Re^{-1}T_0\left(\lambda_0\sD_y + \mu_0\sD_y + \mu_{0,y}\right)\\
\mathcal{L}_{13} &= k_x k_zRe^{-1}T_0(\lambda_0+\mu_0)  \\
\mathcal{L}_{14} &= i k_x\left(\gamma M^2\right)^{-1} T_0^2 \\
\mathcal{L}_{15} &= i k_x\left(\gamma M^2\right)^{-1} - Re^{-1} T_0\left(U_{yy}\mu_{0,T} + U_yT_{0,y}\mu_{0,TT} + U_y\mu_{0,T}\sD_y\right) \\
\mathcal{L}_{21} &= -ik_x Re^{-1}T_0(\lambda_{0,y}+(\lambda_0+\mu_0)\sD_y)\\
\mathcal{L}_{22} &= i k_xU+Re^{-1}T_0\left(\mu_0(k_x^2+k_z^2)-(2\mu_0+\lambda_0)\sD_{yy}-\lambda_{0,y}\sD_y -2\mu_{0,y}\sD_y\right)\\
\mathcal{L}_{23} &= -ik_z Re^{-1}T_0\left(\lambda_{0,y}+(\lambda_0+\mu_0)\sD_y\right)\\
\mathcal{L}_{24} &= \left(\gamma M^2\right)^{-1}T_0(T_{0,y}+T_0\sD_y)\\
\mathcal{L}_{25} &= \left(\gamma M^2\right)^{-1}( T_0\rho_{0,y}+\sD_y) - ik_xRe^{-1} T_0 U_y \mu_{0,T} \\
%\end{align*}
%\begin{align*}
\mathcal{L}_{31} &= k_xk_zRe^{-1}T_0(\mu_0+\lambda_0) \\
\mathcal{L}_{32} &= -ik_z Re^{-1} T_0\left((\mu_0+\lambda_0)\sD_y + \mu_{0,y}\right)\\
\mathcal{L}_{33} &= ik_xU +Re^{-1}T_0\left(\mu_0(k_x^2+k_z^2)-\mu_0\sD_{yy}+k_z^2(\mu_0+\lambda_0)-\mu_{0,y}\sD_y\right)\\
\mathcal{L}_{34} &= ik_z \left(\gamma M^2\right)^{-1}T_0^2 \\
\mathcal{L}_{35} &= ik_z \left(\gamma M^2\right)^{-1} \\
\mathcal{L}_{41} &= ik_z\rho_0 \\
\mathcal{L}_{42} &= \rho_{0,y}+\rho_0\sD_y \\
\mathcal{L}_{43} &= ik_z\rho_0\\
\mathcal{L}_{44} &= ik_x U\\
%\mathcal{L}_{45} &= 0 \\
\mathcal{L}_{51} &= ik_x(\gamma-1)T_0 -2\gamma(\gamma-1)M^2Re^{-1}T_0\mu_0U_y\sD_y \\
\mathcal{L}_{52} &= T_{0,y} + (\gamma-1)T_0\sD_y - 2ik_x\gamma(\gamma-1)M^2Re^{-1}T_0\mu_0U_y \\
\mathcal{L}_{53} &= ik_z(\gamma-1)T_0 \\
%\mathcal{L}_{54} &= 0\\
\mathcal{L}_{55} &= ik_xU-\gamma Re^{-1}\left[Pr^{-1}  T_0\left(2T_{0,y}\mu_{0,T}\sD_y+(T_{0,y})^2\mu_{0,TT}+T_{0,yy}\mu_{0,T} \right.\right.\\
& \left.\quad -\mu_0(k_x^2+k_z^2)+\mu_0\sD_{yy}\right)
 \left.- (\gamma-1)M^2T_0\mu_{0,T}(U_y)^2   \right]
\end{align*}
 Subscripts $y$ and $T$ following commas denote derivatives with respect to wall-normal position and mean temperature respectively, and subscript 0 denotes a mean quantity. Here $\sD$ denotes a derivative operator. $\lambda$ and $\mu$ denote the first and second coefficients of viscosity, related by Stokes' assumption. $\mu$ is assumed to vary with temperature via Sutherland's formula. $\gamma$ represents the ratio of specific heats. The mean states (which vary with Mach number) can be obtained by a simple iteration procedure. $Re$, $M$ and $Pr$ denote the dimensionless Reynolds, Mach, and Prandtl numbers, respectively. 
 
 The corresponding entries to the compressible resolvent operator $\mathcal{H}_{comp}$ are obtained by numerically computing $\left(-i\omega  + \mathcal{L}_{comp}\right)^{-1}$. The fact that we have 23 nonzero blocks of $\mathcal{L}_{comp}$ (compared with 3 for the incompressible equivalent in velocity-vorticity form) makes it infeasible to explicitly show the the block components $\mathcal{H}_{ij}$ in terms of $\mathcal{L}_{ij}$.

%%       \citet{<label>} ==> Lamport (1994)
%%       \citep{<label>} ==> (Lamport, 1994)
%%
%Example citation, See \citet{lamport94}.





% %% For authoryear reference style

% \bibitem[Lamport(1994)]{lamport94}
% Leslie Lamport,
% \textit{\LaTeX: a document preparation system},
% Addison Wesley, Massachusetts,
% 2nd edition,
% 1994.

% \end{thebibliography}

%\section*{References}
\bibliographystyle{elsarticle-harv} 
\bibliography{master.bib}

\begin{thebibliography}{68}
\expandafter\ifx\csname natexlab\endcsname\relax\def\natexlab#1{#1}\fi
\providecommand{\url}[1]{\texttt{#1}}
\providecommand{\href}[2]{#2}
\providecommand{\path}[1]{#1}
\providecommand{\DOIprefix}{doi:}
\providecommand{\ArXivprefix}{arXiv:}
\providecommand{\URLprefix}{URL: }
\providecommand{\Pubmedprefix}{pmid:}
\providecommand{\doi}[1]{\href{http://dx.doi.org/#1}{\path{#1}}}
\providecommand{\Pubmed}[1]{\href{pmid:#1}{\path{#1}}}
\providecommand{\bibinfo}[2]{#2}
\ifx\xfnm\relax \def\xfnm[#1]{\unskip,\space#1}\fi
%Type = Article
\bibitem[{Abreu et~al.(2020)Abreu, Cavalieri, Schlatter, Vinuesa and
  Henningson}]{abreu2020spectral}
\bibinfo{author}{Abreu, L.I.}, \bibinfo{author}{Cavalieri, A.V.G.},
  \bibinfo{author}{Schlatter, P.}, \bibinfo{author}{Vinuesa, R.},
  \bibinfo{author}{Henningson, D.S.}, \bibinfo{year}{2020}.
\newblock \bibinfo{title}{Spectral proper orthogonal decomposition and
  resolvent analysis of near-wall coherent structures in turbulent pipe flows}.
\newblock \bibinfo{journal}{Journal of Fluid Mechanics} \bibinfo{volume}{900}.
%Type = Article
\bibitem[{Bae et~al.(2020)Bae, Dawson and McKeon}]{bae2020resolvent}
\bibinfo{author}{Bae, H.J.}, \bibinfo{author}{Dawson, S.T.M.},
  \bibinfo{author}{McKeon, B.J.}, \bibinfo{year}{2020}.
\newblock \bibinfo{title}{Resolvent-based study of compressibility effects on
  supersonic turbulent boundary layers}.
\newblock \bibinfo{journal}{Journal of Fluid Mechanics} \bibinfo{volume}{883},
  \bibinfo{pages}{A29}.
%Type = Article
\bibitem[{Bhattacharjee et~al.(2024)Bhattacharjee, Mushtaq, Seiler and
  Hemati}]{bhattacharjee2024structured}
\bibinfo{author}{Bhattacharjee, D.}, \bibinfo{author}{Mushtaq, T.},
  \bibinfo{author}{Seiler, P.}, \bibinfo{author}{Hemati, M.S.},
  \bibinfo{year}{2024}.
\newblock \bibinfo{title}{Structured input-output modeling and robust stability
  analysis of compressible flows}.
\newblock \bibinfo{journal}{arXiv preprint arXiv:2407.14986} .
%Type = Inproceedings
\bibitem[{Bhattacharjee et~al.(2023)Bhattacharjee, Mushtaq, Seiler and
  Hemati}]{bhattacharjee2023structured}
\bibinfo{author}{Bhattacharjee, D.}, \bibinfo{author}{Mushtaq, T.},
  \bibinfo{author}{Seiler, P.J.}, \bibinfo{author}{Hemati, M.S.},
  \bibinfo{year}{2023}.
\newblock \bibinfo{title}{Structured input-output analysis of compressible
  plane {Couette} flow}, in: \bibinfo{booktitle}{AIAA SCITECH 2023 Forum}, p.
  \bibinfo{pages}{1984}.
%Type = Article
\bibitem[{Brunton et~al.(2016)Brunton, Proctor and Kutz}]{brunton2016sindy}
\bibinfo{author}{Brunton, S.L.}, \bibinfo{author}{Proctor, J.L.},
  \bibinfo{author}{Kutz, J.N.}, \bibinfo{year}{2016}.
\newblock \bibinfo{title}{Discovering governing equations from data by sparse
  identification of nonlinear dynamical systems}.
\newblock \bibinfo{journal}{Proceedings of the National Academy of Sciences}
  \bibinfo{volume}{113}, \bibinfo{pages}{3932--3937}.
%Type = Article
\bibitem[{Brunton et~al.(2015)Brunton, Proctor, Tu and Kutz}]{Brunton2015jcd}
\bibinfo{author}{Brunton, S.L.}, \bibinfo{author}{Proctor, J.L.},
  \bibinfo{author}{Tu, J.H.}, \bibinfo{author}{Kutz, J.N.},
  \bibinfo{year}{2015}.
\newblock \bibinfo{title}{Compressed sensing and dynamic mode decomposition}.
\newblock \bibinfo{journal}{Journal of Computational Dynamics}
  \bibinfo{volume}{2}, \bibinfo{pages}{165--191}.
%Type = Article
\bibitem[{Callaham et~al.(2021)Callaham, Koch, Brunton, Kutz and
  Brunton}]{callaham2021learning}
\bibinfo{author}{Callaham, J.L.}, \bibinfo{author}{Koch, J.V.},
  \bibinfo{author}{Brunton, B.W.}, \bibinfo{author}{Kutz, J.N.},
  \bibinfo{author}{Brunton, S.L.}, \bibinfo{year}{2021}.
\newblock \bibinfo{title}{Learning dominant physical processes with data-driven
  balance models}.
\newblock \bibinfo{journal}{Nature Communications} \bibinfo{volume}{12},
  \bibinfo{pages}{1--10}.
%Type = Article
\bibitem[{Cand{\`e}s et~al.(2006)Cand{\`e}s, Romberg and
  Tao}]{candes2006robust}
\bibinfo{author}{Cand{\`e}s, E.J.}, \bibinfo{author}{Romberg, J.},
  \bibinfo{author}{Tao, T.}, \bibinfo{year}{2006}.
\newblock \bibinfo{title}{Robust uncertainty principles: Exact signal
  reconstruction from highly incomplete frequency information}.
\newblock \bibinfo{journal}{IEEE Transactions on Information Theory}
  \bibinfo{volume}{52}, \bibinfo{pages}{489--509}.
%Type = Article
\bibitem[{Cand{\`e}s and Wakin(2008)}]{candes2008introduction}
\bibinfo{author}{Cand{\`e}s, E.J.}, \bibinfo{author}{Wakin, M.B.},
  \bibinfo{year}{2008}.
\newblock \bibinfo{title}{An introduction to compressive sampling}.
\newblock \bibinfo{journal}{IEEE Signal Processing Magazine}
  \bibinfo{volume}{25}, \bibinfo{pages}{21--30}.
%Type = Article
\bibitem[{Candon et~al.(2024)Candon, Hale, Delgado-Guti{\'e}rrez, Marzocca and
  Balajewicz}]{candon2024optimal}
\bibinfo{author}{Candon, M.}, \bibinfo{author}{Hale, E.},
  \bibinfo{author}{Delgado-Guti{\'e}rrez, A.}, \bibinfo{author}{Marzocca, P.},
  \bibinfo{author}{Balajewicz, M.}, \bibinfo{year}{2024}.
\newblock \bibinfo{title}{Optimal sparsity in nonlinear nonparametric reduced
  order models for transonic aeroelastic systems}.
\newblock \bibinfo{journal}{AIAA Journal} \bibinfo{volume}{62},
  \bibinfo{pages}{3841--3856}.
%Type = Article
\bibitem[{Chu(1965)}]{chu1965energy}
\bibinfo{author}{Chu, B.T.}, \bibinfo{year}{1965}.
\newblock \bibinfo{title}{On the energy transfer to small disturbances in fluid
  flow (part i)}.
\newblock \bibinfo{journal}{Acta Mechanica} \bibinfo{volume}{1},
  \bibinfo{pages}{215--234}.
%Type = Article
\bibitem[{Dawson and McKeon(2019a)}]{dawson2019shape}
\bibinfo{author}{Dawson, S.T.M.}, \bibinfo{author}{McKeon, B.J.},
  \bibinfo{year}{2019}a.
\newblock \bibinfo{title}{On the shape of resolvent modes in wall-bounded
  turbulence}.
\newblock \bibinfo{journal}{Journal of Fluid Mechanics} \bibinfo{volume}{877},
  \bibinfo{pages}{682--716}.
%Type = Inproceedings
\bibitem[{Dawson and McKeon(2019b)}]{dawson2019couette}
\bibinfo{author}{Dawson, S.T.M.}, \bibinfo{author}{McKeon, B.J.},
  \bibinfo{year}{2019}b.
\newblock \bibinfo{title}{Studying the effects of compressibility in planar
  {C}ouette flow using resolvent analysis}, in: \bibinfo{booktitle}{AIAA
  Scitech 2019 Forum}, p. \bibinfo{pages}{2139}.
%Type = Article
\bibitem[{Dawson and McKeon(2020)}]{dawson2020prediction}
\bibinfo{author}{Dawson, S.T.M.}, \bibinfo{author}{McKeon, B.J.},
  \bibinfo{year}{2020}.
\newblock \bibinfo{title}{Prediction of resolvent mode shapes in supersonic
  turbulent boundary layers}.
\newblock \bibinfo{journal}{International Journal of Heat and Fluid Flow}
  \bibinfo{volume}{85}, \bibinfo{pages}{108677}.
%Type = Article
\bibitem[{Dieci and Papini(2001)}]{dieci2001conditioning}
\bibinfo{author}{Dieci, L.}, \bibinfo{author}{Papini, A.},
  \bibinfo{year}{2001}.
\newblock \bibinfo{title}{Conditioning of the exponential of a block triangular
  matrix}.
\newblock \bibinfo{journal}{Numerical Algorithms} \bibinfo{volume}{28},
  \bibinfo{pages}{137--150}.
%Type = Article
\bibitem[{Duan et~al.(2011)Duan, Beekman and Mart{\'\i}n}]{duan2011direct}
\bibinfo{author}{Duan, L.}, \bibinfo{author}{Beekman, I.},
  \bibinfo{author}{Mart{\'\i}n, M.P.}, \bibinfo{year}{2011}.
\newblock \bibinfo{title}{Direct numerical simulation of hypersonic turbulent
  boundary layers. {P}art 3. effect of {M}ach number}.
\newblock \bibinfo{journal}{Journal of Fluid Mechanics} \bibinfo{volume}{672},
  \bibinfo{pages}{245--267}.
%Type = Article
\bibitem[{Duck et~al.(1994)Duck, Erlebacher and Hussaini}]{duck1994couette}
\bibinfo{author}{Duck, P.W.}, \bibinfo{author}{Erlebacher, G.},
  \bibinfo{author}{Hussaini, M.Y.}, \bibinfo{year}{1994}.
\newblock \bibinfo{title}{On the linear stability of compressible plane couette
  flow}.
\newblock \bibinfo{journal}{Journal of Fluid Mechanics} \bibinfo{volume}{258},
  \bibinfo{pages}{131--165}.
%Type = Article
\bibitem[{Eldar et~al.(2010)Eldar, Kuppinger and Bolcskei}]{eldar2010block}
\bibinfo{author}{Eldar, Y.C.}, \bibinfo{author}{Kuppinger, P.},
  \bibinfo{author}{Bolcskei, H.}, \bibinfo{year}{2010}.
\newblock \bibinfo{title}{Block-sparse signals: Uncertainty relations and
  efficient recovery}.
\newblock \bibinfo{journal}{IEEE Transactions on Signal Processing}
  \bibinfo{volume}{58}, \bibinfo{pages}{3042--3054}.
%Type = Article
\bibitem[{Farghadan et~al.(2023)Farghadan, Martini and
  Towne}]{farghadan2023scalable}
\bibinfo{author}{Farghadan, A.}, \bibinfo{author}{Martini, E.},
  \bibinfo{author}{Towne, A.}, \bibinfo{year}{2023}.
\newblock \bibinfo{title}{Scalable resolvent analysis for three-dimensional
  flows}.
\newblock \bibinfo{journal}{arXiv preprint arXiv:2309.04617} .
%Type = Article
\bibitem[{Foures et~al.(2013)Foures, Caulfield and
  Schmid}]{foures2013localization}
\bibinfo{author}{Foures, D.P.G.}, \bibinfo{author}{Caulfield, C.P.},
  \bibinfo{author}{Schmid, P.J.}, \bibinfo{year}{2013}.
\newblock \bibinfo{title}{Localization of flow structures using $\infty$-norm
  optimization}.
\newblock \bibinfo{journal}{Journal of Fluid Mechanics} \bibinfo{volume}{729},
  \bibinfo{pages}{672--701}.
%Type = Article
\bibitem[{Ganapathisubramani et~al.(2006)Ganapathisubramani, Clemens and
  Dolling}]{ganapathisubramani2006large}
\bibinfo{author}{Ganapathisubramani, B.}, \bibinfo{author}{Clemens, N.T.},
  \bibinfo{author}{Dolling, D.}, \bibinfo{year}{2006}.
\newblock \bibinfo{title}{Large-scale motions in a supersonic turbulent
  boundary layer}.
\newblock \bibinfo{journal}{Journal of Fluid Mechanics} \bibinfo{volume}{556},
  \bibinfo{pages}{271--282}.
%Type = Incollection
\bibitem[{Grant and Boyd(2008)}]{gb08}
\bibinfo{author}{Grant, M.C.}, \bibinfo{author}{Boyd, S.P.},
  \bibinfo{year}{2008}.
\newblock \bibinfo{title}{Graph implementations for nonsmooth convex programs},
  in: \bibinfo{editor}{Blondel, V.}, \bibinfo{editor}{Boyd, S.},
  \bibinfo{editor}{Kimura, H.} (Eds.), \bibinfo{booktitle}{Recent Advances in
  Learning and Control}. \bibinfo{publisher}{Springer-Verlag Limited}. Lecture
  Notes in Control and Information Sciences, pp. \bibinfo{pages}{95--110}.
%Type = Misc
\bibitem[{Grant and Boyd(2014)}]{grant2014cvx}
\bibinfo{author}{Grant, M.C.}, \bibinfo{author}{Boyd, S.P.},
  \bibinfo{year}{2014}.
\newblock \bibinfo{title}{{CVX}: {M}atlab software for disciplined convex
  programming, version 2.1}.
%Type = Article
\bibitem[{Hutchins and Marusic(2007)}]{hutchins2007evidence}
\bibinfo{author}{Hutchins, N.}, \bibinfo{author}{Marusic, I.},
  \bibinfo{year}{2007}.
\newblock \bibinfo{title}{Evidence of very long meandering features in the
  logarithmic region of turbulent boundary layers}.
\newblock \bibinfo{journal}{Journal of Fluid Mechanics} \bibinfo{volume}{579},
  \bibinfo{pages}{1--28}.
%Type = Article
\bibitem[{Hwang and Cossu(2010)}]{hwang2010linear}
\bibinfo{author}{Hwang, Y.}, \bibinfo{author}{Cossu, C.}, \bibinfo{year}{2010}.
\newblock \bibinfo{title}{Linear non-normal energy amplification of harmonic
  and stochastic forcing in the turbulent channel flow}.
\newblock \bibinfo{journal}{Journal of Fluid Mechanics} \bibinfo{volume}{664},
  \bibinfo{pages}{51--73}.
%Type = Article
\bibitem[{Illingworth(2020)}]{illingworth2020streamwise}
\bibinfo{author}{Illingworth, S.J.}, \bibinfo{year}{2020}.
\newblock \bibinfo{title}{Streamwise-constant large-scale structures in
  {C}ouette and {P}oiseuille flows}.
\newblock \bibinfo{journal}{Journal of Fluid Mechanics} \bibinfo{volume}{889}.
%Type = Article
\bibitem[{Jovanovi{\'c}(2021)}]{jovanovic2021bypass}
\bibinfo{author}{Jovanovi{\'c}, M.R.}, \bibinfo{year}{2021}.
\newblock \bibinfo{title}{From bypass transition to flow control and
  data-driven turbulence modeling: An input--output viewpoint}.
\newblock \bibinfo{journal}{Annual Review of Fluid Mechanics}
  \bibinfo{volume}{53}, \bibinfo{pages}{311--345}.
%Type = Article
\bibitem[{Jovanovi{\'c} and Bamieh(2005)}]{jovanovic2005componentwise}
\bibinfo{author}{Jovanovi{\'c}, M.R.}, \bibinfo{author}{Bamieh, B.},
  \bibinfo{year}{2005}.
\newblock \bibinfo{title}{Componentwise energy amplification in channel flows}.
\newblock \bibinfo{journal}{Journal of Fluid Mechanics} \bibinfo{volume}{534},
  \bibinfo{pages}{145--183}.
%Type = Article
\bibitem[{Jovanovi{\'c} et~al.(2014)Jovanovi{\'c}, Schmid and
  Nichols}]{jovanovic2014dmdsp}
\bibinfo{author}{Jovanovi{\'c}, M.R.}, \bibinfo{author}{Schmid, P.J.},
  \bibinfo{author}{Nichols, J.W.}, \bibinfo{year}{2014}.
\newblock \bibinfo{title}{Sparsity-promoting dynamic mode decomposition}.
\newblock \bibinfo{journal}{Physics of Fluids} \bibinfo{volume}{26},
  \bibinfo{pages}{--}.
%Type = Article
\bibitem[{Kaiser et~al.(2018)Kaiser, Kutz and Brunton}]{kaiser2018sparse}
\bibinfo{author}{Kaiser, E.}, \bibinfo{author}{Kutz, J.N.},
  \bibinfo{author}{Brunton, S.L.}, \bibinfo{year}{2018}.
\newblock \bibinfo{title}{Sparse identification of nonlinear dynamics for model
  predictive control in the low-data limit}.
\newblock \bibinfo{journal}{Proceedings of the Royal Society A}
  \bibinfo{volume}{474}, \bibinfo{pages}{20180335}.
%Type = Article
\bibitem[{Kline et~al.(1967)Kline, Reynolds, Schraub and
  Runstadler}]{kline1967structure}
\bibinfo{author}{Kline, S.J.}, \bibinfo{author}{Reynolds, W.C.},
  \bibinfo{author}{Schraub, F.A.}, \bibinfo{author}{Runstadler, P.W.},
  \bibinfo{year}{1967}.
\newblock \bibinfo{title}{The structure of turbulent boundary layers}.
\newblock \bibinfo{journal}{Journal of Fluid Mechanics} \bibinfo{volume}{30},
  \bibinfo{pages}{741--773}.
%Type = Article
\bibitem[{Kramer et~al.(2017)Kramer, Grover, Boufounos, Nabi and
  Benosman}]{kramer2017sparse}
\bibinfo{author}{Kramer, B.}, \bibinfo{author}{Grover, P.},
  \bibinfo{author}{Boufounos, P.}, \bibinfo{author}{Nabi, S.},
  \bibinfo{author}{Benosman, M.}, \bibinfo{year}{2017}.
\newblock \bibinfo{title}{Sparse sensing and dmd-based identification of flow
  regimes and bifurcations in complex flows}.
\newblock \bibinfo{journal}{SIAM Journal on Applied Dynamical Systems}
  \bibinfo{volume}{16}, \bibinfo{pages}{1164--1196}.
%Type = Article
\bibitem[{Landahl(1975)}]{landahl1975wave}
\bibinfo{author}{Landahl, M.T.}, \bibinfo{year}{1975}.
\newblock \bibinfo{title}{Wave breakdown and turbulence}.
\newblock \bibinfo{journal}{SIAM Journal on Applied Mathematics}
  \bibinfo{volume}{28}, \bibinfo{pages}{735--756}.
%Type = Article
\bibitem[{Loiseau and Brunton(2018)}]{loiseau2018constrained}
\bibinfo{author}{Loiseau, J.C.}, \bibinfo{author}{Brunton, S.L.},
  \bibinfo{year}{2018}.
\newblock \bibinfo{title}{Constrained sparse {G}alerkin regression}.
\newblock \bibinfo{journal}{Journal of Fluid Mechanics} \bibinfo{volume}{838},
  \bibinfo{pages}{42--67}.
%Type = Article
\bibitem[{Lopez-Doriga et~al.(2024)Lopez-Doriga, Ballouz, Bae and
  Dawson}]{lopez2024sparse}
\bibinfo{author}{Lopez-Doriga, B.}, \bibinfo{author}{Ballouz, E.},
  \bibinfo{author}{Bae, H.J.}, \bibinfo{author}{Dawson, S.T.},
  \bibinfo{year}{2024}.
\newblock \bibinfo{title}{Sparse space–time resolvent analysis for
  statistically stationary and time-varying flows}.
\newblock \bibinfo{journal}{Journal of Fluid Mechanics} \bibinfo{volume}{999},
  \bibinfo{pages}{A87}.
%Type = Article
\bibitem[{Malik et~al.(2006)Malik, Alam and Dey}]{malik2006nonmodal}
\bibinfo{author}{Malik, M.}, \bibinfo{author}{Alam, M.}, \bibinfo{author}{Dey,
  J.}, \bibinfo{year}{2006}.
\newblock \bibinfo{title}{Nonmodal energy growth and optimal perturbations in
  compressible plane {C}ouette flow}.
\newblock \bibinfo{journal}{Physics of Fluids} \bibinfo{volume}{18},
  \bibinfo{pages}{034103}.
%Type = Article
\bibitem[{Martini et~al.(2021)Martini, Rodr{\'\i}guez, Towne and
  Cavalieri}]{martini2021efficient}
\bibinfo{author}{Martini, E.}, \bibinfo{author}{Rodr{\'\i}guez, D.},
  \bibinfo{author}{Towne, A.}, \bibinfo{author}{Cavalieri, A.V.G.},
  \bibinfo{year}{2021}.
\newblock \bibinfo{title}{Efficient computation of global resolvent modes}.
\newblock \bibinfo{journal}{Journal of Fluid Mechanics} \bibinfo{volume}{919}.
%Type = Article
\bibitem[{McKeon and Sharma(2010)}]{mckeon2010resolvent}
\bibinfo{author}{McKeon, B.J.}, \bibinfo{author}{Sharma, A.S.},
  \bibinfo{year}{2010}.
\newblock \bibinfo{title}{A critical-layer framework for turbulent pipe flow}.
\newblock \bibinfo{journal}{Journal of Fluid Mechanics} \bibinfo{volume}{658},
  \bibinfo{pages}{336--382}.
%Type = Article
\bibitem[{Moarref et~al.(2013)Moarref, Sharma, Tropp and
  McKeon}]{moarref2013channels}
\bibinfo{author}{Moarref, R.}, \bibinfo{author}{Sharma, A.S.},
  \bibinfo{author}{Tropp, J.A.}, \bibinfo{author}{McKeon, B.J.},
  \bibinfo{year}{2013}.
\newblock \bibinfo{title}{Model-based scaling of the streamwise energy density
  in high-{R}eynolds-number turbulent channels}.
\newblock \bibinfo{journal}{Journal of Fluid Mechanics} \bibinfo{volume}{734},
  \bibinfo{pages}{275--316}.
%Type = Article
\bibitem[{Morkovin(1962)}]{morkovin1962hypothesis}
\bibinfo{author}{Morkovin, M.V.}, \bibinfo{year}{1962}.
\newblock \bibinfo{title}{Effects of compressibility on turbulent flows}.
\newblock \bibinfo{journal}{M{\'e}canique de la Turbulence}
  \bibinfo{volume}{367}, \bibinfo{pages}{380}.
%Type = Article
\bibitem[{Mushtaq and Hemati(2024)}]{mushtaq2024identifying}
\bibinfo{author}{Mushtaq, T.}, \bibinfo{author}{Hemati, M.S.},
  \bibinfo{year}{2024}.
\newblock \bibinfo{title}{Identifying spatially-localized instability
  mechanisms using sparse optimization}.
\newblock \bibinfo{journal}{arXiv preprint arXiv:2411.05617} .
%Type = Article
\bibitem[{Nogueira et~al.(2021)Nogueira, Morra, Martini, Cavalieri and
  Henningson}]{nogueira2021forcing}
\bibinfo{author}{Nogueira, P.A.S.}, \bibinfo{author}{Morra, P.},
  \bibinfo{author}{Martini, E.}, \bibinfo{author}{Cavalieri, A.V.G.},
  \bibinfo{author}{Henningson, D.S.}, \bibinfo{year}{2021}.
\newblock \bibinfo{title}{Forcing statistics in resolvent analysis: application
  in minimal turbulent {C}ouette flow}.
\newblock \bibinfo{journal}{Journal of Fluid Mechanics} \bibinfo{volume}{908}.
%Type = Inproceedings
\bibitem[{Orr(1907)}]{orr1907stability}
\bibinfo{author}{Orr, W.M.}, \bibinfo{year}{1907}.
\newblock \bibinfo{title}{The stability or instability of the steady motions of
  a perfect liquid and of a viscous liquid. {P}art {II}: A viscous liquid}, in:
  \bibinfo{booktitle}{Proceedings of the Royal Irish Academy. Section A:
  Mathematical and Physical Sciences}, \bibinfo{organization}{JSTOR}. pp.
  \bibinfo{pages}{69--138}.
%Type = Article
\bibitem[{Pan et~al.(2021)Pan, Arnold-Medabalimi and
  Duraisamy}]{pan2021sparsity}
\bibinfo{author}{Pan, S.}, \bibinfo{author}{Arnold-Medabalimi, N.},
  \bibinfo{author}{Duraisamy, K.}, \bibinfo{year}{2021}.
\newblock \bibinfo{title}{Sparsity-promoting algorithms for the discovery of
  informative {Koopman}-invariant subspaces}.
\newblock \bibinfo{journal}{Journal of Fluid Mechanics} \bibinfo{volume}{917},
  \bibinfo{pages}{A18}.
%Type = Article
\bibitem[{Pickering et~al.(2021)Pickering, Rigas, Schmidt, Sipp and
  Colonius}]{pickering2021optimal}
\bibinfo{author}{Pickering, E.}, \bibinfo{author}{Rigas, G.},
  \bibinfo{author}{Schmidt, O.T.}, \bibinfo{author}{Sipp, D.},
  \bibinfo{author}{Colonius, T.}, \bibinfo{year}{2021}.
\newblock \bibinfo{title}{Optimal eddy viscosity for resolvent-based models of
  coherent structures in turbulent jets}.
\newblock \bibinfo{journal}{Journal of Fluid Mechanics} \bibinfo{volume}{917}.
%Type = Article
\bibitem[{Reynolds and Tiederman(1967)}]{reynolds1967stability}
\bibinfo{author}{Reynolds, W.C.}, \bibinfo{author}{Tiederman, W.G.},
  \bibinfo{year}{1967}.
\newblock \bibinfo{title}{Stability of turbulent channel flow, with application
  to {M}alkus's theory}.
\newblock \bibinfo{journal}{Journal of Fluid Mechanics} \bibinfo{volume}{27},
  \bibinfo{pages}{253--272}.
%Type = Article
\bibitem[{Ribeiro et~al.(2020)Ribeiro, Yeh and Taira}]{ribeiro2020randomized}
\bibinfo{author}{Ribeiro, J.H.M.}, \bibinfo{author}{Yeh, C.A.},
  \bibinfo{author}{Taira, K.}, \bibinfo{year}{2020}.
\newblock \bibinfo{title}{Randomized resolvent analysis}.
\newblock \bibinfo{journal}{Physical Review Fluids} \bibinfo{volume}{5},
  \bibinfo{pages}{033902}.
%Type = Article
\bibitem[{Rosenberg and McKeon(2019)}]{rosenberg2018efficient}
\bibinfo{author}{Rosenberg, K.}, \bibinfo{author}{McKeon, B.J.},
  \bibinfo{year}{2019}.
\newblock \bibinfo{title}{Efficient representation of exact coherent states of
  the {N}avier--{S}tokes equations using resolvent analysis}.
\newblock \bibinfo{journal}{Fluid Dynamics Research} \bibinfo{volume}{51},
  \bibinfo{pages}{011401}.
%Type = Article
\bibitem[{Rubini et~al.(2020)Rubini, Lasagna and Da~Ronch}]{rubini2020l1}
\bibinfo{author}{Rubini, R.}, \bibinfo{author}{Lasagna, D.},
  \bibinfo{author}{Da~Ronch, A.}, \bibinfo{year}{2020}.
\newblock \bibinfo{title}{The $l_1$-based sparsification of energy interactions
  in unsteady lid-driven cavity flow}.
\newblock \bibinfo{journal}{Journal of Fluid Mechanics} \bibinfo{volume}{905}.
%Type = Article
\bibitem[{Rubini et~al.(2022)Rubini, Lasagna and Da~Ronch}]{rubini2022priori}
\bibinfo{author}{Rubini, R.}, \bibinfo{author}{Lasagna, D.},
  \bibinfo{author}{Da~Ronch, A.}, \bibinfo{year}{2022}.
\newblock \bibinfo{title}{A priori sparsification of {Galerkin} models}.
\newblock \bibinfo{journal}{Journal of Fluid Mechanics} \bibinfo{volume}{941},
  \bibinfo{pages}{A43}.
%Type = Article
\bibitem[{Samimy et~al.(1994)Samimy, Arnette and
  Elliott}]{samimy1994streamwise}
\bibinfo{author}{Samimy, M.}, \bibinfo{author}{Arnette, S.},
  \bibinfo{author}{Elliott, G.S.}, \bibinfo{year}{1994}.
\newblock \bibinfo{title}{Streamwise structures in a turbulent supersonic
  boundary layer}.
\newblock \bibinfo{journal}{Physics of fluids} \bibinfo{volume}{6},
  \bibinfo{pages}{1081--1083}.
%Type = Article
\bibitem[{Schmid(2010)}]{Schmid:2010}
\bibinfo{author}{Schmid, P.J.}, \bibinfo{year}{2010}.
\newblock \bibinfo{title}{Dynamic mode decomposition of numerical and
  experimental data}.
\newblock \bibinfo{journal}{Journal of Fluid Mechanics} \bibinfo{volume}{656},
  \bibinfo{pages}{5--28}.
%Type = Book
\bibitem[{Schmid and Henningson(2012)}]{schmid2012book}
\bibinfo{author}{Schmid, P.J.}, \bibinfo{author}{Henningson, D.S.},
  \bibinfo{year}{2012}.
\newblock \bibinfo{title}{Stability and transition in shear flows}.
\newblock \bibinfo{publisher}{Springer Science \& Business Media}.
%Type = Article
\bibitem[{Skene et~al.(2022)Skene, Yeh, Schmid and
  Taira}]{skene2022sparsifying}
\bibinfo{author}{Skene, C.S.}, \bibinfo{author}{Yeh, C.A.},
  \bibinfo{author}{Schmid, P.J.}, \bibinfo{author}{Taira, K.},
  \bibinfo{year}{2022}.
\newblock \bibinfo{title}{Sparsifying the resolvent forcing mode via
  gradient-based optimisation}.
\newblock \bibinfo{journal}{Journal of Fluid Mechanics} \bibinfo{volume}{944}.
%Type = Article
\bibitem[{Smith and Smits(1995)}]{smith1995visualization}
\bibinfo{author}{Smith, M.}, \bibinfo{author}{Smits, A.}, \bibinfo{year}{1995}.
\newblock \bibinfo{title}{Visualization of the structure of supersonic
  turbulent boundary layers}.
\newblock \bibinfo{journal}{Experiments in Fluids} \bibinfo{volume}{18},
  \bibinfo{pages}{288--302}.
%Type = Book
\bibitem[{Smits and Dussauge(2006)}]{smits2006supersonic}
\bibinfo{author}{Smits, A.J.}, \bibinfo{author}{Dussauge, J.P.},
  \bibinfo{year}{2006}.
\newblock \bibinfo{title}{Turbulent shear layers in supersonic flow}.
\newblock \bibinfo{publisher}{Springer Science \& Business Media}.
%Type = Article
\bibitem[{Smits et~al.(1989)Smits, Spina, Alving, Smith, Fernando and
  Donovan}]{smits1989comparison}
\bibinfo{author}{Smits, A.J.}, \bibinfo{author}{Spina, E.F.},
  \bibinfo{author}{Alving, A.E.}, \bibinfo{author}{Smith, R.W.},
  \bibinfo{author}{Fernando, E.M.}, \bibinfo{author}{Donovan, J.F.},
  \bibinfo{year}{1989}.
\newblock \bibinfo{title}{A comparison of the turbulence structure of subsonic
  and supersonic boundary layers}.
\newblock \bibinfo{journal}{Physics of Fluids A: Fluid Dynamics}
  \bibinfo{volume}{1}, \bibinfo{pages}{1865--1875}.
%Type = Article
\bibitem[{Symon et~al.(2023)Symon, Madhusudanan, Illingworth and
  Marusic}]{symon2023use}
\bibinfo{author}{Symon, S.}, \bibinfo{author}{Madhusudanan, A.},
  \bibinfo{author}{Illingworth, S.J.}, \bibinfo{author}{Marusic, I.},
  \bibinfo{year}{2023}.
\newblock \bibinfo{title}{Use of eddy viscosity in resolvent analysis of
  turbulent channel flow}.
\newblock \bibinfo{journal}{Physical Review Fluids} \bibinfo{volume}{8},
  \bibinfo{pages}{064601}.
%Type = Article
\bibitem[{Tissot et~al.(2021)Tissot, Cavalieri and
  M{\'e}min}]{tissot2021stochastic}
\bibinfo{author}{Tissot, G.}, \bibinfo{author}{Cavalieri, A.V.G.},
  \bibinfo{author}{M{\'e}min, E.}, \bibinfo{year}{2021}.
\newblock \bibinfo{title}{Stochastic linear modes in a turbulent channel flow}.
\newblock \bibinfo{journal}{Journal of Fluid Mechanics} \bibinfo{volume}{912}.
%Type = Article
\bibitem[{Toh et~al.(1999)Toh, Todd and T{\"u}t{\"u}nc{\"u}}]{toh1999sdpt3}
\bibinfo{author}{Toh, K.C.}, \bibinfo{author}{Todd, M.J.},
  \bibinfo{author}{T{\"u}t{\"u}nc{\"u}, R.H.}, \bibinfo{year}{1999}.
\newblock \bibinfo{title}{{SDPT3—a M}atlab software package for semidefinite
  programming, version 1.3}.
\newblock \bibinfo{journal}{Optimization methods and software}
  \bibinfo{volume}{11}, \bibinfo{pages}{545--581}.
%Type = Article
\bibitem[{Towne et~al.(2018)Towne, Schmidt and Colonius}]{towne2018spectral}
\bibinfo{author}{Towne, A.}, \bibinfo{author}{Schmidt, O.T.},
  \bibinfo{author}{Colonius, T.}, \bibinfo{year}{2018}.
\newblock \bibinfo{title}{Spectral proper orthogonal decomposition and its
  relationship to dynamic mode decomposition and resolvent analysis}.
\newblock \bibinfo{journal}{Journal of Fluid Mechanics} \bibinfo{volume}{847},
  \bibinfo{pages}{821--867}.
%Type = Book
\bibitem[{Trefethen(2000)}]{Trefethen:2000}
\bibinfo{author}{Trefethen, L.N.}, \bibinfo{year}{2000}.
\newblock \bibinfo{title}{Spectral Methods in {MATLAB}}.
%Type = Article
\bibitem[{Trefethen et~al.(1993)Trefethen, Trefethen, Reddy and
  Driscoll}]{trefethen1993science}
\bibinfo{author}{Trefethen, L.N.}, \bibinfo{author}{Trefethen, A.E.},
  \bibinfo{author}{Reddy, S.C.}, \bibinfo{author}{Driscoll, T.A.},
  \bibinfo{year}{1993}.
\newblock \bibinfo{title}{Hydrodynamic stability without eigenvalues}.
\newblock \bibinfo{journal}{Science} \bibinfo{volume}{261},
  \bibinfo{pages}{578--584}.
%Type = Article
\bibitem[{Tu et~al.(2014)Tu, Rowley, Kutz and Shang}]{tu2014compressed}
\bibinfo{author}{Tu, J.H.}, \bibinfo{author}{Rowley, C.W.},
  \bibinfo{author}{Kutz, J.N.}, \bibinfo{author}{Shang, J.K.},
  \bibinfo{year}{2014}.
\newblock \bibinfo{title}{Spectral analysis of fluid flows using
  sub-{Nyquist}-rate {PIV} data}.
\newblock \bibinfo{journal}{Experiments in Fluids} \bibinfo{volume}{55},
  \bibinfo{pages}{1--13}.
%Type = Article
\bibitem[{T{\"u}t{\"u}nc{\"u} et~al.(2003)T{\"u}t{\"u}nc{\"u}, Toh and
  Todd}]{tutuncu2003solving}
\bibinfo{author}{T{\"u}t{\"u}nc{\"u}, R.H.}, \bibinfo{author}{Toh, K.C.},
  \bibinfo{author}{Todd, M.J.}, \bibinfo{year}{2003}.
\newblock \bibinfo{title}{Solving semidefinite-quadratic-linear programs using
  {SDPT3}}.
\newblock \bibinfo{journal}{Mathematical programming} \bibinfo{volume}{95},
  \bibinfo{pages}{189--217}.
%Type = Article
\bibitem[{Weideman and Reddy(2000)}]{weideman2000matlab}
\bibinfo{author}{Weideman, J.A.}, \bibinfo{author}{Reddy, S.C.},
  \bibinfo{year}{2000}.
\newblock \bibinfo{title}{A {M}atlab differentiation matrix suite}.
\newblock \bibinfo{journal}{ACM Transactions on Mathematical Software (TOMS)}
  \bibinfo{volume}{26}, \bibinfo{pages}{465--519}.
%Type = Article
\bibitem[{Wen et~al.(2010)Wen, Goldfarb and Yin}]{wen2010alternating}
\bibinfo{author}{Wen, Z.}, \bibinfo{author}{Goldfarb, D.},
  \bibinfo{author}{Yin, W.}, \bibinfo{year}{2010}.
\newblock \bibinfo{title}{Alternating direction augmented lagrangian methods
  for semidefinite programming}.
\newblock \bibinfo{journal}{Mathematical Programming Computation}
  \bibinfo{volume}{2}, \bibinfo{pages}{203--230}.
%Type = Article
\bibitem[{Zhou et~al.(1999)Zhou, Adrian, Balachandar and
  Kendall}]{zhou1999mechanisms}
\bibinfo{author}{Zhou, J.}, \bibinfo{author}{Adrian, R.J.},
  \bibinfo{author}{Balachandar, S.}, \bibinfo{author}{Kendall, T.M.},
  \bibinfo{year}{1999}.
\newblock \bibinfo{title}{Mechanisms for generating coherent packets of hairpin
  vortices in channel flow}.
\newblock \bibinfo{journal}{Journal of Fluid Mechanics} \bibinfo{volume}{387},
  \bibinfo{pages}{353--396}.

\end{thebibliography}

\end{document}